\pdfoutput=1
\documentclass[pra,aps,12pt,floatfix,notitlepage,nofootinbib,superscriptaddress,longbibliography]{revtex4-2}

\newcommand{\be}{\begin{equation}}
\newcommand{\ee}{\end{equation}}
\newcommand{\bea}{\begin{eqnarray}}
\newcommand{\eea}{\end{eqnarray}}

\newcommand{\meanv}[1]{\langle #1 \rangle}

\newcommand{\meanvlr}[1]{\left\langle #1 \right\rangle}

\newcommand{\bb}[1]{\left( #1 \right)}

\newcommand{\bbcro}[1]{\left[ #1 \right]}

\newcommand{\ii}{\textrm{i}}
\newcommand{\dd}{\textrm{d}}
\newcommand{\eee}{\textrm{e}}

\newcommand{\qq}{\textbf{q}}

\newcommand{\pp}{\textbf{p}}

\newcommand{\ab}{\overline{a}}

\newcommand{\ket}[1]{|#1\rangle}
\newcommand{\bra}[1]{\langle #1|}

\newcommand{\kF}{k_{\rm F}}

\newcommand{\TF}{T_{\rm F}}

\newcommand{\pF}{p_{\rm F}}
\newcommand{\vF}{v_{\rm F}}
\newcommand{\EF}{\epsilon_{\rm F}}
\newcommand{\FS}{\ket{{\rm FS}}}

\newcommand{\upa}{\uparrow}
\newcommand{\dwa}{\downarrow}

\newcommand{\eqqref}[1]{Eq.~\eqref{#1}}
\newcommand{\eqqrefs}[2]{Eqs.~\eqref{#1}--\eqref{#2}}

\usepackage{tikz}
\usetikzlibrary{shapes.geometric, positioning, shadings, arrows.meta}
\usetikzlibrary{calc,patterns,decorations.pathmorphing,decorations.markings}

\tikzset{->-/.style={decoration={
  markings,
  mark=at position .5 with {\arrow{>}}},postaction={decorate}}}
\tikzset{-<-/.style={decoration={
  markings,
  mark=at position .5 with {\arrow{<}}},postaction={decorate}}}
\tikzset{phantom->-/.style={decoration={
  markings,
  mark=at position .5 with {\arrow[scale=2]{>}}},postaction={decorate}}}

\tikzset
  {
    ,my bubble/.style = 
      {
        ,draw=#1!70
        ,fill=#1!10
        ,ellipse
        ,inner sep=1pt
        ,minimum width=2em
        ,minimum height=2em
        ,align=center
      }
    ,my end/.style =
      {
        ,draw=#1!70
        ,top color=#1!10
        ,bottom color=#1!50
        ,minimum height=6em
        ,text width=6em
        ,inner sep=0pt
        ,align=center
      }
    ,my arrow/.style =
      {
        ,>=Stealth
        ,->
        ,draw=black
      }
  }
  
\tikzset{serpent/.style={decoration={snake},postaction={decorate}}}

\usepackage{amsmath}
\usepackage{feynmp-auto}
\usepackage[compat=1.1.0]{tikz-feynman}
\usepackage[mathscr]{euscript}
 \let\mathscr\relax 
\usepackage[scr]{rsfso}
\usepackage{amsthm}
\usepackage{cancel}
\usepackage{amssymb}
\usepackage[normalem]{ulem}
\usepackage{soul}
\soulregister\cite7
\soulregister\ref7

\newcommand{\grp}{\textbf{p}}
\newcommand{\gq}{\textbf{q}}

\newcommand{\barre}[1]{\overline{#1}}

\usetikzlibrary{shapes.geometric, positioning, shadings, arrows.meta}
\usetikzlibrary{calc,patterns,decorations.pathmorphing,decorations.markings}
\tikzset{->-/.style={decoration={
  markings,
  mark=at position .5 with {\arrow{>}}},postaction={decorate}}}
\tikzset{-<-/.style={decoration={
  markings,
  mark=at position .5 with {\arrow{<}}},postaction={decorate}}}
\tikzset{phantom->-/.style={decoration={
  markings,
  mark=at position .5 with {\arrow[scale=2]{>}}},postaction={decorate}}}

\tikzset
  {
    ,my bubble/.style = 
      {
        ,draw=#1!70
        ,fill=#1!10
        ,ellipse
        ,inner sep=1pt
        ,minimum width=2em
        ,minimum height=2em
        ,align=center
      }
    ,my end/.style =
      {
        ,draw=#1!70
        ,top color=#1!10
        ,bottom color=#1!50
        ,minimum height=6em
        ,text width=6em
        ,inner sep=0pt
        ,align=center
      }
    ,my arrow/.style =
      {
        ,>=Stealth
        ,->
        ,draw=black
      }
  }
  
\tikzset{serpent/.style={decoration={snake},postaction={decorate}}}

\counterwithin*{paragraph}{subsection}
\usepackage[breaklinks=true]{hyperref}
\hypersetup{
    colorlinks,
    linkcolor={red!50!black},
    citecolor={blue!90},
    urlcolor={blue!80!black}
}
\usepackage{xcite}
\usepackage{xr-hyper}

\renewcommand{\theequation}{II.\arabic{equation}}
\begin{document}

\title{A low-energy effective Hamiltonian for Landau quasiparticles:\\ II. Application to the contact Fermi gas}

\author{Pierre-Louis Taillat }
\email{pierre-louis.taillat@sorbonne-universite.fr}
\affiliation{Laboratoire de Physique Théorique de la Matière Condensée, Sorbonne Université, CNRS, 75005, Paris, France}
\author{Hadrien Kurkjian}
\email{hadrien.kurkjian@cnrs.fr}
\affiliation{Laboratoire de Physique Théorique de la Matière Condensée, Sorbonne Université, CNRS, 75005, Paris, France}

\date{\today}
\begin{abstract}
This article follows up on Ref.~\cite{VolumeI}, in which we developed a new renormalization scheme to construct a quantized theory of Fermi liquids.
 Here, we apply this formalism to a low-temperature atomic Fermi gas where the short-range interactions are fully parametrized by the s-wave scattering length $a$.
 We benchmark our renormalized theory by recovering known perturbative results on the static properties of the Fermi gas, such as the Lee-Huang-Yang expansion
 of the equation of state, the Galitskii expansion of the momentum distribution, and the Gor'kov-Melik Barkhudarov preexponential
 correction to the critical temperature.
We then turn to the transport dynamics and demonstrate the presence of a preexponential correction to the speed of zero sound when including corrections of second order in $a$. Finally, we develop an efficient numerical method to solve the transport equation exactly, and we apply it to study the crossover from the collisionless to the hydrodynamic regime in the density and polarisation response functions.
\end{abstract}
\maketitle

\vspace{\baselineskip}

\tableofcontents

\section*{Introduction}

Following the construction of the low-energy effective Hamiltonian of a generic
Fermi liquid in the first part \cite{VolumeI} of this publication, we apply here the
formalism to a two-component gas of ultracold fermionic atoms. In ultracold gases, the bare atom-atom interactions,
arising from the van der Waals forces, are characterized by a range $r_0$
much shorter than both the interparticle distance $\approx \kF^{-1}$
and the thermal de Broglie wavelength $\approx(mT)^{-1/2}$. In this cold and dilute regime
of collisions, the true interatomic potential can be replaced by an effective one,
adjusted to correctly describe the low-energy scattering events 
which are dominated by $\upa$-$\dwa$
scattering in the s-wave channel. The effective atom-atom potential (not to be confused
with the quasiparticle potential) can then be chosen as a contact potential,
whose coupling constant $g_0$ reproduces the s-wave scattering length $a$.

\begin{figure}[htb]
\begin{center}
\includegraphics[width=0.8\columnwidth]{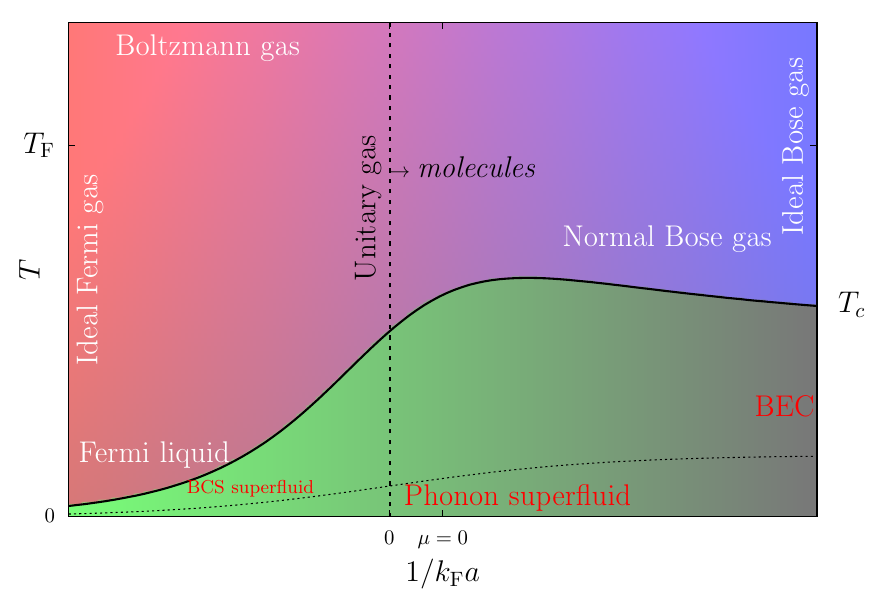}
\end{center}
\caption{\label{fig:diagphase} Equilibrium phase diagram of the spin-$1/2$ Fermi gas with contact interactions. The normal
and superfluid phases are separated by the critical temperature $T_c$ (solid line). The various
limiting regimes of the normal and superfluid phase are shown in white and red respectively.}
\end{figure}

The equilibrium state of this contact Fermi gas depends 
solely on three thermodynamic parameters, the density $\rho$, the temperature $T$ and 
the scattering length $a$, usually combined via the Fermi units in the two dimensionless 
parameters $T/\TF$ and $1/\kF a$. The corresponding phase diagram is shown 
in Fig.~\ref{fig:diagphase}. It contains a normal phase at high temperature ($T>T_c$)
and a superfluid phase at low temperature. As a function of the interaction parameter
$1/\kF a$ a smooth transition occurs from a fermionic regime for $1/\kF a\to-\infty$
to a bosonic regime for $1/\kF a\to+\infty$. 
This so-called BEC-BCS crossover
is not restricted to the superfluid phase and also describes the transition from a normal
Fermi gas to a normal Bose gas at $T>T_c$.
The progressive appearance  of a bosonic behavior can be seen as a consequence
of the existence of a two-body bound state (a dimer) of the contact potential for $a>0$.
This dimer becomes infinitely bound when $a\to0^+$ which ensures that the internal
fermionic degrees of freedom become irrelevant for the many-body behavior.

The contact Fermi gas is widely regarded as the archetype of the neutral Fermi fluids:
the remarkable simplicity of its microscopic physics provides an ideal setting 
to explore the fermionic many-body problem in its purest form.
For this reason, it has attracted a lot of experimental and theoretical attention
since its first realization two decades ago \cite{Jin2003,Zwerger2012}. Its equilibrium
properties, including its phase diagram and its equation-of-state,
are well understood by now, both theoretically, experimentally, and numerically.

The attention has thus shifted from the equilibrium to the dynamical properties,
a subject on which there remains difficult open questions.
In the strongly-interacting regime ($\kF |a|\approx 1$) and at intermediate temperatures
($T/\TF\approx 1$), the evolution of the contact Fermi gas
is a strongly-correlated problem (even at long wavelengths and low energies),
in the sense that it cannot be captured by a one-body phase-space distribution
obeying a kinetic equation. Higher-order correlation functions,
which one may organise into a BBGKY hierarchy \cite{Bonitz1998,Kira2015,eindhoven},
may not be truncated without committing an uncontrolled error.

\begin{figure}[htb]
\begin{center}
\includegraphics[width=0.9\columnwidth]{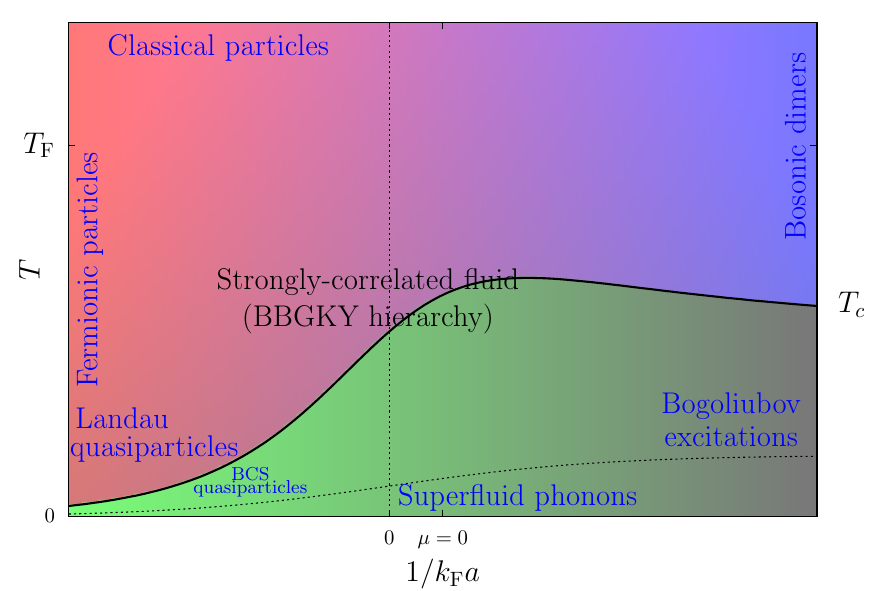}
\end{center}
\caption{The phase diagram 
showing (in blue) the microscopic degrees
of freedom that support a kinetic equation in various
limiting regimes. Away from these limiting regimes (that is for $T\approx \TF$ and $\kF a\approx 1$)
transport is a strongly-correlated phenomenon. \label{diagtransport}}
\end{figure}

Fortunately, kinetics equations are recovered in several limiting regimes (shown in Fig.~\ref{diagtransport}),
which gives us access to controlled results, and crucial benchmarks
for numerical or experimental results that access the most difficult region of
the phase diagram \cite{Thomas2008,Grimm2008,Zwierlein2011,Zwierlein2011Universal,Kohl2012,Kohl2013,Thywissen2014,Thywissen2015,Thywissen2018,Thomas2022,yaleexp}.
Much effort have been devoted for instance to derive the transport properties
at unitarity ($|a|=+\infty$) \cite{Schafer2007,Smith2007,Zwerger2011},
starting from the virial expansion of the kinetic equation of the high-temperature gas \cite{Schafer2010,Nishida2019,Enss2019,Hofmann2020,Enss2023b}.
Similarly, reliable predictions are available in the very low temperature regime ($T\ll T_c$)
where the superfluid behaves as a (mainly collisionless) gas of phonons \cite{Schafer2007,Zwerger2011,annalen,landaukhalat,amorbf,Tsimokha2026}.

There is however another strategy to approach the strongly-interacting regime, which starts
from the low-temperature ideal Fermi gas (at $a\to0^-$) \cite{Vichi2000,Nikuni2009}, and expands the transport properties
in powers of $\kF a$. Despite its experimental relevance \cite{Zwierlein2011Universal,yaleexp}
and theoretical challenge, this strategy has been somewhat overlooked in the recent literature,
and we fill the knowledge gap here. 

Obtaining non-perturbative expressions of the effective functions $f_{\sigma\sigma'}$, $g_{\sigma\sigma'}$
and $\mathcal{A}_{\sigma\sigma'}$ in function of the scattering length $a$
is a very difficult task, and controlled results are likely accessible only
to heavy numerical approaches. Here, we stick to a perturbative treatment,
valid in the limit $a\to0^-$, and perform a perturbative calculation to second-order in 
$\kF a$. In this convenient regime, the unitary quasiparticle transformation
can be performed straight through, without a need to reduce $\Lambda$ continuously.
The flow equation remain however useful to extract the renormalised effective parameters from the
perturbative expressions, of \textit{e.g.} the pair interaction function $g_{\sigma\sigma'}$.

In Section \ref{sec:applicationGFC}, we apply the Schrieffer-Wolff
unitary transformation perturbatively to the contact potential. This results
in an expansion of the quasiparticle parameters up to subleading order in $\kF a$. 
We benchmark our method by recovering the static properties of the contact 
Fermi gas, such as the Lee-Huang-Yang equation of state \cite{Yang1957} or
the particle momentum distribution. In our effective picture, the Gor'kov Melik-Barkhudarov 
correction to $T_c$ follows simply from the second-order corrections to the
renormalized pair interaction $G_{\sigma\sigma'}$.
We finally verify with the contact potential the Bethe-Salpeter relations derived in Section \ref{sec:flot} on the forward
and frontal collision amplitudes.

In Section~\ref{sec:collless}, we turn to the dynamical properties,
which are far less understood than the static ones, even in the pertubative regime. 
Since the hydrodynamic regime is treated in Ref.~\cite{devvisco}, 
we focus on the density and spin responses in the collisionless regime.
A major result is the presence of a preexponential factor in the expression of the
zero sound velocity $c_0$ in terms of $\kF a$, reminiscent of the GMB correction to $T_c$. 
This multiplies the deviation between $c_0$ and $\vF$
by $\exp(6)$ for the density mode and $\exp(-2)$ for the spin mode. 
We further investigate the role of collisions on zero sound and show that they lead to a collisional damping proportional to $1/\tau$, 
where $\tau$ is the mean collision time. 
This damping is universal in the sense that its dependence on $k_{\rm F}a$ enters only through $\tau$. 

Finally, in Section~\ref{sec:collless2} we present a numerical method that solves the transport equation exactly throughout the crossover 
from the hydrodynamic to the collisionless regime. In
the contact Fermi gas, we distinguish between a shallow collisionless regime $\vF/(c_0 - \vF)>\omega\tau\gg 1$
where the collisional damping of zero sound is sufficient to dilute the resonance in the quasiparticle-hole continuum, and
a deep collisionless regime $\omega\tau>\vF/(c_0 - \vF)$, where the resonance detaches from the continuum.

To guide the reader through the notations a table of symbols is available in Appendix \ref{formulaire} of Ref.\cite{VolumeI}.

\section{Effective description of a low-temperature Fermi gas with contact interactions}
\label{sec:applicationGFC}

\subsection{Contact potential}
\label{sec:contactpotential}

The Hamiltonian of the Fermi gas with contact interactions
is a special case of the generic form introduced in section \ref{sec:HRG}:
\begin{eqnarray}
\hat{H}_0&=&\sum_{\pp\in\mathcal{D},\sigma} \omega_\pp \hat a_{\pp\sigma}^\dagger \hat a_{\pp\sigma}, \qquad \omega_\pp=\frac{p^2}{2m}\label{H0contact}\\
\hat{V}&=&\frac{g_0}{L^3}\sum_{\pp_1,\pp_2,\pp_3,\pp_4\in
\mathcal{D}} \delta_{\pp_1+\pp_2}^{\pp_3+\pp_4} 
\hat a_{\pp_1\upa}^\dagger \hat a_{\pp_2\dwa}^\dagger \hat a_{\pp_3\dwa} \hat a_{\pp_4\upa} \label{Vcontact}
\end{eqnarray} 

To regularize the UV divergences inherent to the contact potential, we have discretized the real space \cite{Varenna,CastinLesHouches2004} into a cubic lattice of step $l$, thereby restricting the set of momenta $\pp$ to 
$\mathcal{D}=(2\pi \mathbb{Z}/L)^3\ \cap\ [-\pi/l,\pi/l[^3$. Solving the two-body problem, we express the bare coupling consant $g_0$ in terms of $a$ through the Lippmann-Schwinger equation
\begin{equation}
\frac{1}{g_0}=\frac{1}{g}-\int_{[-\pi/l,\pi/l[^3}\frac{\dd^3 p}{(2\pi)^3}\frac{m}{p^2} \label{gg0}
\end{equation}
where 
\be
g=\frac{4\pi a}{m}
\ee
Rather than sending the lattice spacing $l$ to 0 at fixed 
$a$, as is usually done when treating the BEC-BCS crossover
non-perturbatively, 
%his is regularized by eliminating $g_0$ in favor of $a$. 
here we regularize the divergences perturbatively by expanding the Lippman-Schwinger relation \eqref{gg0} for $a\to0$ at fixed $l$
\cite{Varenna}:
\begin{equation}
g_0=g+g^2 \mathcal{P}\underset{[-\pi/l,\pi/l)^3}\int\frac{\dd^3 p_a \dd^3 p_b}{(2\pi)^3}\frac{\delta(\pp_1+\pp_2-\pp_a-\pp_b)}{(p_a^2+p_b^2-p_1^2-p_2^2)/2m}
+O(g^3) \label{gg0exp}
\end{equation}
The integral on the right-hand side shows a UV divergence
which combines with a counter divergence in the second-order terms
of the many-body quantities (see Ref.~\cite{LandauLipschitzVol9} and \eqqref{Enps}, \eqref{epsilonpkFa} below), 
allowing to send finally $l\to0$.
The integral in \eqref{gg0exp} is a rewriting of the one appearing in \eqqref{gg0}:
the difference between the two integrals is independent of $\pp_1$, $\pp_2$ and vanishes in the continuous limit $l\to0$.

Finally, remark that the contact potential respects the SU(2) symmetry \eqqref{SU2}.
\subsection{Perturbative Schrieffer-Wolff decomposition}
\label{sec:perturbative}

When $\hat V$ is controlled by a small parameter ($g_0$ in our case), one can construct perturbatively
the operator $\hat S$
and the effective Hamiltonian $\hat H_{\rm eff}$ introduced in
Sec.~\ref{unitaire}:
\begin{equation}
    \hat{S} = \hat{S}_1 + \hat{S}_2 + \ldots \ \text{ where } \ \hat{S}_1 = {O}(V),\ \hat{S}_2 = {O}(V^2), \ldots \label{expS}
\end{equation}
Expanding to first order in $\hat V$ condition (\ref{PHP}) on the band-diagonality of $\hat H_{\rm eff}$, we obtain
\be
\hat Q_\Lambda \bb{\hat V +[\hat H_0,\hat S_1]}\hat{P}_\Lambda=0 \label{PmS1Pn}
%\hat{P}_m \bb{[\hat V,S_1] +[\hat H_0,\hat S_2]+\frac{1}{2}\bbcro{[\hat H_0,\hat S_1],\hat S_1}}\hat{P}_n=0
\ee
This relation provides the elements of $\hat S_1$ in the unperturbed basis:
\be
{}_0\bra{f} \hat S_1\ket{i}_0=
\begin{cases}
\frac{{}_0\bra{f} \hat V \ket{i}_0}{E_i^0-E_f^0} &\text{ if } |E_i^0-E_f^0|>\Lambda \\
\ 0 &\text{ if } |E_i^0-E_f^0|<\Lambda
\end{cases}
\ee
One can also express the restriction of $\hat S_1$ to a given energy shell $\hat P_\Lambda(E)$ 
in terms of the unperturbed resolvent $\hat G_0(E)=1/(E-\hat H_0)$:
\be
\hat S_1 \hat P_\Lambda(E)= \hat Q_\Lambda(E) \bbcro{ \hat G_0(E)\hat V} \hat P_\Lambda(E)
\ee
Note however that $\hat S$, unlike $\hat G_0(E)$, does not depend on energy and allows
to (block) diagonalize the whole spectrum, not just the vicinity of a particular energy level.

Using the perturbative expression of $\hat S$ and the Baker-Campbell-Haussdorf
expansion \eqref{HeffBCH} of $\hat H_{\rm eff}$, we obtain\footnote{
To simplify to double commutator $\hat P_\Lambda\bbcro{[\hat{H}_0,\hat{S}_1],\hat{S}_1}\hat P_\Lambda=-\hat P_\Lambda\bbcro{\hat V,\hat{S}_1}\hat P_\Lambda$, we have used  \eqqref{PmS1Pn} and $\hat P_\Lambda[\hat H_0,\hat S_1+\hat S_2]\hat P_\Lambda=0$.} a perturbative expression of the effective Hamiltonian:
\be
\hat P_\Lambda \hat H_{\rm eff}\hat P_\Lambda=\hat P_\Lambda \bb{\hat H_0+\hat V+\frac{1}{2}[\hat V,\hat S_1]+O(V^3)}\hat P_\Lambda
\ee

\subsection{Effective Hamiltonian of the contact Fermi gas}
%\subsection{Expansion of the quasiparticle annihilation operator}

Applied to the contact potential, \eqqref{PmS1Pn}  provides the expression of $\hat S$ to leading order in $g$:
\be
\hat S_1=\frac{g}{L^3}\sum_{\pp_\alpha\pp_\beta \pp_\gamma{\pp_\delta}\in\mathcal{D}} \hat a_{\pp_\alpha\upa}^\dagger \hat a_{\pp_\beta\dwa}^\dagger \hat a_{\pp_\gamma\dwa} \hat a_{{\pp_\delta}\upa} \delta_{\pp_\alpha+\pp_\beta}^{\pp_\gamma+{\pp_\delta}}\ \mathcal{P}_{\Lambda}\bb{\frac{1}{\omega_{\pp_\gamma}+\omega_{{\pp_\delta}}-\omega_{\pp_\alpha}-\omega_{\pp_\beta}}} \label{S1contact}
\ee
Here, the $\Lambda$-principal part
\be
\mathcal{P}_{\Lambda}\bb{\frac{1}{E}}=\begin{cases}1/E\text{ if }|E|>\Lambda \\ 0 \text{ else} \end{cases}
\ee
originates in the projectors $\hat P_{\Lambda}$ and prevents the denominators from vanishing.
As a first result, we derive from the definition \eqqref{gammaformelle} an explicit expression the quasiparticle annihilation operator
\begin{multline}
\hat\gamma_{\pp\upa}=\hat{a}_{\pp\upa} + [\hat{S}_1,\hat{a}_{\pp\upa}]+\ldots=\hat a_{\pp\upa}\\+\frac{g}{L^3}\sum_{\pp_\beta\pp_\gamma{\pp_\delta}\in\mathcal{D}}  \delta_{\pp+\pp_\beta}^{\pp_\gamma+{\pp_\delta}}\ \mathcal{P}_{\Lambda}\bb{\frac{1}{\omega_{\pp_\gamma}+\omega_{{\pp_\delta}}-\omega_{\pp}-\omega_{\pp_\beta}}}\hat a_{\pp_\beta\dwa}^\dagger \hat a_{\pp_\gamma\dwa} \hat a_{{\pp_\delta}\upa}+O(a^2) \label{gamma}
\end{multline}
We view \eqqref{gamma} as a rigorous formulation of
the standard first-order picture of the spin $\upa$ quasiparticle as a cloud of spin $\dwa$ particle surrounding a spin $\upa$ particle.
It is reminiscent of the Chevy Ansatz for polarons \cite{Chevy2006}.

%\subsection{Expression of the Hamiltonian in terms of $\hat\gamma$}
When expressing the Hamiltonian $\hat H$ as an effective
Hamiltonian written in terms 
of the operators $\hat \gamma$ (as discussed in Section \ref{sec:HRG}), 
it is convenient to compute the unrestricted Hamiltonian $\hat H'$
\be
%\hat P_{\Lambda,\gamma} \hat H \hat P_{\Lambda,\gamma}=\hat P_{\Lambda,\gamma} \bb{\hat H'+O(g^3)}\hat P_{\Lambda,\gamma} \\
\hat H'=\hat H_{0,\gamma}+\hat V_\gamma+\frac{1}{2}[\hat V_\gamma,S_{1,\gamma}]  \label{Hp} 
\ee
where we recall that $\hat H_{0,\gamma}$, $\hat V_\gamma$ and $\hat S_{1,\gamma}$ follow
from Eqs.~\eqref{H0contact}, \eqref{Vcontact} and \eqref{S1contact} through the replacement $\hat a\to\hat \gamma$.
We then implement the quasi-resonance condition coming from the projectors  at the end of the calculation.
We have
\begin{multline}
\hat{ H}'=\sum_{\pp\sigma}\omega_{\pp} \gamma_{\pp\sigma}^\dagger \gamma_{\pp\sigma}+\frac{g_0}{L^3} \sum_{\pp_\alpha\pp_\beta\pp_\gamma{\pp_\delta}\in\mathcal{D}}\delta_{\pp_\alpha+\pp_\beta}^{\pp_\gamma+{\pp_\delta}} 
\hat \gamma_{\pp_\alpha\upa}^\dagger \hat \gamma_{\pp_\beta\dwa}^\dagger \hat \gamma_{\pp_\gamma\dwa} \hat a_{{\pp_\delta}\upa}
+\frac{1}{2}\bb{\frac{g_0}{L^3}}^2\\\times\sum_{\substack{\pp_\alpha\pp_\beta\pp_\gamma{\pp_\delta}\in\mathcal{D}\\\pp_a\pp_b\pp_c\pp_d\in\mathcal{D}}} \delta_{\pp_\alpha+\pp_\beta}^{\pp_\gamma+{\pp_\delta}} \delta_{\pp_a+\pp_b}^{\pp_c+\pp_d} \mathcal{P}_{\Lambda}\bb{\frac{1}{\omega_{\pp_a}+\omega_{\pp_b}-\omega_{\pp_c}-\omega_{\pp_d}}}  \bbcro{\hat \gamma_{\pp_\alpha\upa}^\dagger \hat \gamma_{\pp_\beta\dwa}^\dagger \hat \gamma_{\pp_\gamma\dwa} \hat \gamma_{{\pp_\delta}\upa},\hat \gamma_{\pp_d\upa}^\dagger \hat \gamma_{\pp_c\dwa}^\dagger \hat \gamma_{\pp_b\dwa} \hat \gamma_{\pp_a\upa}}  
\label{Hpcomm}
\end{multline}
This Hamiltonian truncated to second order in $\hat V$ is thus sextic 
in $\hat\gamma$, with up to $3\leftrightarrow3$, as discussed in Section \ref{collisionamplitudes} (see note \ref{note:collisionsn}).

Rather than computing the commutator in \eqqref{Hpcomm}, we directly linearize about the 
quasiparticle Fermi sea $\ket{{\rm FS}}$ by injecting the expansion $\hat \gamma^\dagger \hat \gamma=\bra{\rm FS}\hat \gamma^\dagger \hat \gamma\FS+\delta(\hat \gamma^\dagger \hat \gamma)$ \eqqref{deltagammagamma}.
This method preserves the detail balance structure of the fermionic occupation numbers
which follows from the commutator.
We note that there is no ambiguity in how to pair the operators $\hat\gamma$ before injecting the decomposition. 
The $\Lambda$-principal part $\mathcal{P}_{\Lambda}$ in \eqqref{Hpcomm} guarantees that $a\neq d$, $b\neq c$ and thus
\be
(\hat \gamma_{\pp_\alpha\upa}^\dagger \hat \gamma_{\pp_\beta\dwa}^\dagger \hat \gamma_{\pp_\gamma\dwa} \hat \gamma_{{\pp_\delta}\upa})(\hat \gamma_{\pp_d\upa}^\dagger \hat \gamma_{\pp_c\dwa}^\dagger \hat \gamma_{\pp_b\dwa} \hat \gamma_{\pp_a\upa}) =
(\hat \gamma_{\pp_\alpha\upa}^\dagger \hat \gamma_{\pp_a\upa}) (\hat \gamma_{\pp_\beta\dwa}^\dagger \hat \gamma_{\pp_b\dwa}) (\hat \gamma_{\pp_\gamma\dwa}  \hat \gamma_{\pp_c\dwa}^\dagger ) (\hat \gamma_{{\pp_\delta}\upa}\hat \gamma_{\pp_d\upa}^\dagger)
\ee
We obtain an effective Hamiltonian in the form of \eqqref{HFS}:
\begin{multline}
\hat H'=E_{\rm FS}+\sum_{\pp\sigma}\epsilon_{\pp}\delta \hat n_{\pp\sigma} 
+\frac{1}{2L^3}\! \! \!\! \sum_{\substack{\pp_\alpha\pp_\beta\pp_\gamma\pp_\delta\in\mathcal{D}\\\sigma\sigma'=\upa\dwa}}
\delta_{\pp_\alpha+\pp_\beta}^{\pp_\gamma+\pp_{\delta}} {\mathcal{B}}_{\sigma\sigma'}'(\pp_\alpha\pp_\beta|\pp_\gamma\pp_{\delta}) \delta(\hat\gamma_{\pp_\alpha\sigma}^\dagger \hat\gamma_{\pp_{\delta}\sigma})\delta(\hat\gamma_{\pp_\beta\sigma'}^\dagger \hat\gamma_{\pp_\gamma\sigma'})\\+O(\delta(\hat\gamma^\dagger \hat \gamma)^3) \label{HFScontact}
\end{multline}

The effective parameters $E_{\rm FS}$, $\epsilon_{\pp}$ and $\mathcal{B}$ are evaluated 
here perturbatively to second-order in $g$.
For the constant term, \textit{i.e.} the energy of the quasiparticle Fermi sea, we have
\begin{multline}
E_{\rm FS}=\sum_{\pp\in\mathcal{D},\sigma}\omega_{\pp} n_{p}^0+\frac{g}{L^3} \sum_{\pp,\pp'\in\mathcal{D}} n_{\pp}^0 n_{\pp'}^0\\
+\bb{\frac{g}{L^3}}^2 \sum_{\pp_1\pp_2\pp_3\pp_4\in\mathcal{D}}\delta_{\pp_1+\pp_2}^{\pp_3+\pp_4} n_{\pp_1}^0 n_{\pp_2}^0 (\bar n_{\pp_3}^0 \bar n_{\pp_4}^0-1) \mathcal{P}_{\Lambda}\bb{\frac{1}{\omega_{\pp_1}+\omega_{\pp_2}-\omega_{\pp_3}-\omega_{\pp_4}}}+O(a^3) \label{Enps}
\end{multline}
Note that we have used the standard perturbative renormalization procedure \cite{LandauLipschitzVol9,devvisco} to replace $g_0$ by $g$, such that \eqqref{Enps}
is free from UV divergence when $l\to0$.
The explicit of calculation of $E_{\rm FS}(\kF a)$ from \eqqref{Enps} leads to the Lee-Huang-Yang
equation of state to second order in $\kF a$ \cite{LandauLipschitzVol9}.

The eigenenergy  of the quasiparticles is given by
\begin{multline}
\epsilon_{\pp}=\omega_{\pp}+\frac{g}{L^3} \sum_{\pp'\in\mathcal{D}}  n_{\pp'}^0 
+\bb{\frac{g}{L^3}}^2 \sum_{\pp_2\pp_3{\pp_4}\in\mathcal{D}} \delta_{\pp+\pp_2}^{\pp_3+{\pp_4}}\bbcro{n_{\pp_2}^0 (\bar n_{\pp_3}^0 \bar n_{\pp_4}^0-1)+\bar n_{\pp_2}^0  n_{\pp_3}^0  n_{\pp_{\pp_4}}^0} \\\times \mathcal{P}_{\Lambda}\bb{\frac{1}{\omega_{\pp}+\omega_{\pp_2}-\omega_{\pp_3}-\omega_{\pp_4}}} +O(a^3) \label{epsilonpkFa}
\end{multline}
The integrals in this expression were computed analytically by Galitskii \cite{Galitskii1958},
who provided in particular the second-order correction to the effective mass:
\be
\frac{m^\ast}{m}=1+\frac{8 (\kF a)^2}{15\pi^2}(7\text{ln}2-1)+O(a^3)
\ee
Although the low-energy properties are sensitive only to this effective mass,
we note that the expression of $\epsilon_\pp$ in our Hamiltonian renormalization
formalism is not restricted to the vicinity of $\pF$. Unlike the momentum-shell, 
or Wilsonian RG approaches, where the large momenta
are integrated out of the low-energy action, the Hamiltonian RG
removes only the off-shell transitions and conserves the high-momentum degrees of freedom.
The behavior of $\epsilon_\pp$ away from $\pF$ is studied in Refs.~\cite{Galitskii1958,Mahaux1980}.
The curvature $\dd^2\epsilon_\pp/\dd p^2$ in particular determines the bulk viscosity \cite{Brooker1970}.

Finally, the (unconstrained) collision amplitudes $\mathcal{B}'$ are
\begin{multline}
\mathcal{B}_{\sigma\sigma}'(\pp_\alpha\pp_\beta|\pp_\gamma{\pp_\delta}) ={\frac{g^2}{2L^3}} \sum_{\pp_1 \pp_2\in\mathcal{D}}\bbcro{ n_{\pp_1}^0\bar n_{\pp_2}^0-\bar n_{\pp_1}^0n_{\pp_2}^0}\\
\times\bbcro{\mathcal{P}_{\Lambda}\bb{\frac{1}{\omega_{\pp_\alpha}+\omega_{\pp_2}-\omega_{\pp_\gamma}-\omega_{\pp_1}}}\delta_{\pp_1+\pp_\gamma}^{\pp_2+\pp_\alpha}+\mathcal{P}_{\Lambda}\bb{\frac{1}{\omega_{\pp_\beta}+\omega_{\pp_2}-\omega_{\pp_\delta}-\omega_{\pp_1}}}\delta_{\pp_1+\pp_\delta}^{\pp_2+\pp_\beta} }+O(a^3) \label{Bpss}
\end{multline}
\begin{multline}
\mathcal{B}_{\upa\dwa}'(\pp_\alpha\pp_\beta|\pp_\gamma{\pp_\delta}) -\mathcal{B}_{\sigma\sigma}'(\pp_\alpha\pp_\beta|\pp_\gamma{\pp_\delta})={g}+{\frac{g^2}{2L^3}} \sum_{\pp_1\pp_2\in\mathcal{D}}\bbcro{\bar n_{\pp_1}^0\bar n_{\pp_2}^0-n_{\pp_1}^0n_{\pp_2}^0-1}\delta_{\pp_\alpha+\pp_\beta}^{\pp_1+\pp_2}\\\times\bbcro{\mathcal{P}_{\Lambda}\bb{\frac{1}{\omega_{\pp_\alpha}+\omega_{\pp_\beta}-\omega_{\pp_1}-\omega_{\pp_2}}}+\mathcal{P}_{\Lambda}\bb{\frac{1}{\omega_{\pp_\gamma}+\omega_{\pp_\delta}-\omega_{\pp_1}-\omega_{\pp_2}}} }+O(a^3) \label{Bpud}
\end{multline}
Remark that we have symmetrized $\mathcal{B}_{\sigma\sigma}$ towards the full exchange 
$\mathcal{B}_{\sigma\sigma}(\beta,\alpha|\delta,\gamma)=$\newline $\mathcal{B}_{\sigma\sigma}(\alpha,\beta|\gamma,\delta)$, 
and that the function $\mathcal{P}_{\Lambda}$ imposes $\mathcal{B}_{\sigma\sigma}(\pp\pp'|\pp\pp')=0$,
in accordance with the constraints \eqref{contrainteB2}--\eqref{contrainteB1}.
The true amplitude $\mathcal{B}$ are deduced from the unconstrained amplitudes
$\mathcal{B}'$ by reinserting the quasi-resonance condition:
\be
{\mathcal{B}}_{\sigma\sigma'}(\pp_\alpha\pp_\beta|\pp_\gamma\pp_{\delta})= {\mathcal{B}}_{\sigma\sigma'}'(\pp_\alpha\pp_\beta|\pp_\gamma\pp_{\delta}) \Pi_{\Lambda}(\epsilon_{\pp_\alpha}+\epsilon_{\pp_\beta}-\epsilon_{\pp_\gamma}-\epsilon_{\pp_\delta})
\label{Bprime}
\ee

%The standard expression of the Landau interaction function $f_{\sigma\sigma'}$
%in Ref.~\cite{LandauLipschitzVol9} is a special case ($\pp=\pp_\alpha=\pp_\delta$ and $\pp'=\pp_\beta=\pp_\gamma$)
%of \eqqrefs{Bpud}{Bpss}.
%\be
%f_{\sigma\sigma}(\pp,\pp') ={\frac{g^2}{L^3}} \sum_{\pp_1 \pp_2\in\mathcal{D}}\bbcro{ n_{\pp_1}^0\bar n_{\pp_2}^0-\bar n_{\pp_1}^0n_{\pp_2}^0} \delta_{\pp+\pp_2}^{\pp'+\pp_1}
%\mathcal{P}_{\Lambda}\bb{\frac{1}{\omega_{\pp}+\omega_{\pp_2}-\omega_{\pp'}-\omega_{\pp_1}}}+O(g)^3
%\ee
%\begin{multline}
%f_{\upa\dwa}(\pp,\pp') -{f}_{\sigma\sigma}(\pp,\pp')=g+{\frac{g^2}{L^3}} \sum_{\pp_1\pp_2\in\mathcal{D}}\bbcro{\bar n_{\pp_1}^0\bar n_{\pp_2}^0-n_{\pp_1}^0n_{\pp_2}^0-1}\delta_{\pp_\alpha+\pp_\beta}^{\pp_1+\pp_2} \mathcal{P}_{\Lambda}\bb{\frac{1}{\omega_{\pp_\alpha}+\omega_{\pp_\beta}-\omega_{\pp_1}-\omega_{\pp_2}}}  \\+O(g)^3
%\end{multline}

\subsection{Explicit expression of the interaction functions and collision amplitudes}
\label{sec:lambdadep}

From  \eqqrefs{Bpss}{Bpud}, we now evaluate the amplitudes $\mathcal{B}_{\sigma\sigma'}$ on
the Fermi surface ($p_1=p_2=$ $p_3=p_4=\pF$), where they depend only on the angles $\theta_{ij}=(\widehat{\pp_i,\pp_j})$:
\bea
\frac{\mathcal{B}_{\upa\dwa}^\Lambda(\pp_1,\pp_2|\pp_3,\pp_4)}{g}&=&1+\frac{2k_{\rm F}a}{\pi}\bbcro{I_\Lambda(\theta_{12})+J_\Lambda(\theta_{13})}+O(a^2) \label{Bupdw} \\
\frac{\mathcal{B}_{\upa\upa}^\Lambda(\pp_1,\pp_2|\pp_3,\pp_4)}{g}&=&\frac{2k_{\rm F}a}{\pi}{J_\Lambda(\theta_{13})}+O(a^3) \label{Bupup}
\eea
The functions $I_\Lambda$ and $J_\Lambda$ that characterize the crossed $\Lambda$, $\theta$
dependence of $\mathcal{B}$ are depicted on Figs.~\ref{ilambda}--\ref{jlambda}, and explicit expressions are given in Appendix \ref{app:ijlambda}.
\begin{figure}[htb]
\begin{center}
\includegraphics[width=0.7\textwidth]{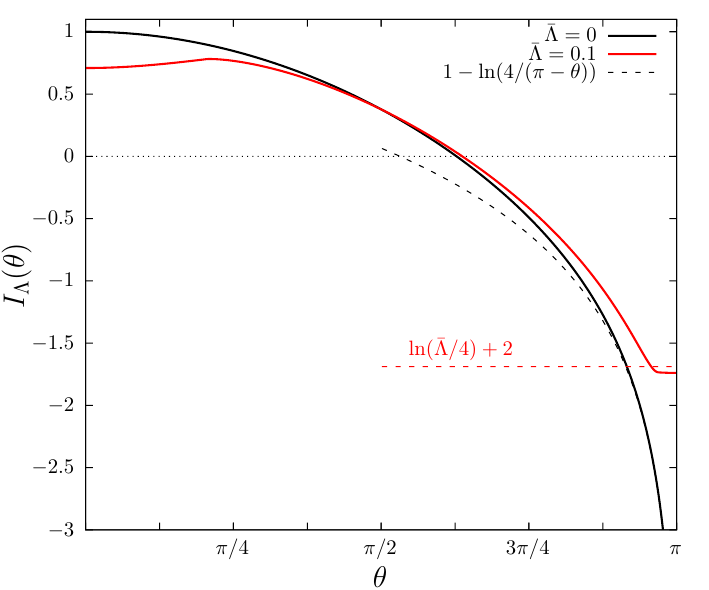}
\end{center}
\caption{\label{ilambda}Angular dependence of the function $I_\Lambda$ appearing in $\mathcal{A}_{\upa\dwa}$ and $f_{\upa\dwa}$. For $\bar\Lambda=0$ (black curve), the function
displays a logarithmic divergence $\sim \text{ln}(\pi-\theta)+1-\text{ln}\,4$ when $\theta\to\pi$ (black dashed curve). For $\bar\Lambda\neq0$ (red curve) the divergence is regularized, and the function saturates at $\text{ln}(\bar\Lambda/4)+2+O(\bar\Lambda)$ in $\theta=\pi$ (red dashed curve).}
\end{figure}

\begin{figure}[htb]
\begin{center}
\includegraphics[width=0.7\textwidth]{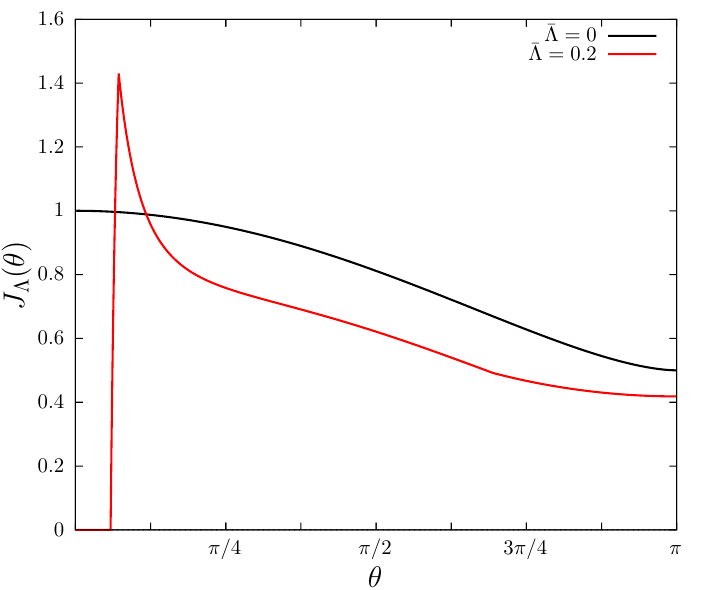}
\end{center}
\caption{\label{jlambda}Angular dependence of the function $J_\Lambda$ appearing in both $\mathcal{A}_{\upa\dwa},$ $f_{\upa\dwa}$ and $\mathcal{A}_{\sigma\sigma},$ $f_{\sigma\sigma}$. As $\bar\Lambda\to0$, the function converges pointwise to $J(\theta)$ (black curve) on $(0,\pi]$. It is however cancelled in an interval of width $\simeq\bar\Lambda$ about $\theta=0$ (red curve).}
\end{figure}

The physical quantities are related to the $\Lambda\to 0$ limit of $I_\Lambda$
and $J_\Lambda$, which does not commute with the small angle limits $\theta\to0,\pi$. 
As discussed in Section \ref{sec:flot}, in those non-analytic channels the collision amplitudes $\mathcal{A}_{\sigma\sigma'}$
is connected to the interaction functions $f_{\sigma\sigma'}$ and $g_{\sigma\sigma'}$. 
%We recall the dimensionless small parameter associated to $\Lambda$
%\be
%\bar\Lambda=\frac{\Lambda}{\vF\pF}
%\ee
When the limit $\bar\Lambda=\Lambda/\vF\pF\to0$ is taken first, the functions converge pointwise in the open interval $(0,\pi)$ to the functions $I$ and $J$ usually
found in this context \cite{LandauLipschitzVol9,devvisco}
\bea
I(\theta)\equiv\text{lim}_{\bar\Lambda\to0} I_\Lambda(\theta) &=&1-\frac{s}{2}\text{ln}\frac{1+s}{1-s}, \qquad s=\sin\frac{\theta}{2}, \ c=\cos\frac{\theta}{2} \label{itheta}\\
J(\theta)\equiv\text{lim}_{\bar\Lambda\to0} J_\Lambda(\theta)&=&\frac{1}{2}\bb{1+\frac{c^2}{2s}\text{ln}\frac{1+s}{1-s}} \label{jtheta}
\eea
While $J$ is a smooth function in $[0,\pi]$, we note a logarithmic divergence in
$I$ when $\theta\to\pi$ (see Fig.~\ref{ilambda}).
As visible in Figs.\ref{ilambda} and \ref{jlambda}, the convergence of $I_\Lambda$ and $J_\Lambda$  to $I$ and $J$  is not uniform: $\bar\Lambda$ regularises
the divergence of $I$ in $\theta=\pi$, and cancels $J$ in a neighborhood of size $\approx\bar\Lambda$ about $\theta=0$. 
Taking now the limit $\theta\to0$ or $\pi$  before $\bar\Lambda\to0$, we have:
\bea
\text{lim}_{\theta\to\pi} I_\Lambda(\theta)&=&\text{ln}\frac{\bar\Lambda}{4}+2+O(\bar\Lambda) \label{limlimI} \\
\text{lim}_{\bar\Lambda\to0} \text{lim}_{\theta\to0} J_\Lambda(\theta)&=&0 \label{limlimJ}
\eea
We recover with these two points of non-uniform convergence the forward ($\theta=0$) and frontal ($\theta=\pi$)
collision channels. The singular behavior of $J_\Lambda(\theta\to0)$ is reminiscent
of the behavior of the 4-point vertex found in Ref.~\cite{Senechal1998} (see Fig.~4 therein).

The collision amplitudes are obtained in the limit $\Lambda\to0$ at non-vanishing $\theta_{ij}$,
which gives\footnote{We recall
that $\mathcal{A}_{\upa\dwa}=\mathcal{B}_{\upa\dwa}$
and $\mathcal{A}_{\sigma\sigma}(\pp_1,\pp_2|$ $\pp_3,\pp_4)$ $=\mathcal{B}_{\sigma\sigma}(\pp_1,\pp_2|\pp_3,\pp_4)-\mathcal{B}_{\sigma\sigma}(\pp_1,\pp_2|\pp_4,\pp_3)$, 
 see Eqs.~\eqref{Bud2} and \eqref{Bss2}}:
\bea
\frac{\mathcal{A}_{\upa\dwa}(\pp_1,\pp_2|\pp_3,\pp_4)}{g}&=&1+\frac{2k_{\rm F}a}{\pi}\bbcro{I(\theta_{12})+J(\theta_{13})}+O(a^2) \label{Aupdw} \\
\frac{\mathcal{A}_{\upa\upa}(\pp_1,\pp_2|\pp_3,\pp_4)}{g}&=&\frac{2k_{\rm F}a}{\pi}\bbcro{J(\theta_{13})-J(\theta_{14})}+O(a^3) \label{Aupup}
\eea
Conversely, the Landau functions $f_{\sigma\sigma'}$ are obtained 
from \eqqrefs{Bupdw}{Bupup} by setting $\theta_{14}=0$  
before letting $\Lambda\to0$:
\begin{eqnarray}
\frac{f_{\upa\dwa}(\pp,\pp')}{g}&=&1+\frac{2k_{\rm F} a}{\pi} \bbcro{I(\theta)+J(\theta)}+O(a^2)\label{fupdw}\\
\frac{f_{\upa\upa}(\pp,\pp')}{g}&=&\frac{2k_{\rm F} a}{\pi} J(\theta)+O(a^2) \label{fupup}
\eea
Here $\theta\equiv(\widehat{\pp,\pp'})$ is equal to both  $\theta_{12}$ and $\theta_{13}$ in \eqqrefs{Bupdw}{Bupup}, corresponding to the forward configuration $\pp=\pp_1=\pp_4$ and $\pp'=\pp_2=\pp_3$.
Similarly,  the pair interaction $g_{\sigma\sigma'}$ (which retains a logarithmic
dependence on $\Lambda$ see Sec.~\ref{BSsec}..c) follows from 
$\theta_{12}=\pi$ before $\Lambda\to0$;
\begin{eqnarray}
\frac{g_{\upa\dwa}(\pp,\pp',\Lambda)}{g}&=&1+\frac{2k_{\rm F} a}{\pi} \bbcro{\text{ln}\frac{\bar\Lambda}{4}+2+J(\theta)}+O(a^2)\label{gupdw}\\
\frac{g_{\sigma\sigma}(\pp,\pp',\Lambda)}{g}&=&\frac{2k_{\rm F} a}{\pi} \bbcro{J(\pi-\theta)-J(\theta)}+O(a^2) \label{gupup}
\eea
Here $\theta\equiv(\widehat{\pp,\pp'})$ is equal to both $\theta_{14}$ and $\pi-\theta_{13}$ in \eqqrefs{Bupdw}{Bupup}, corresponding to the frontal configuration $\pp=\pp_1=-\pp_2$ and $\pp'=\pp_4=-\pp_3$.
Note that the expected logarithmic suppression of $g_{\upa\dwa}(\Lambda)$ appears as a logarithmic \textit{divergence} in \eqqref{gupdw} due to the perturbative expansion.
The logarithmic behavior of $g_{\upa\upa}(\Lambda)$ is even completely absent in the second-order result \eqref{gupup}.

\subsection{Bethe-Salpeter equations}
Comparing the perturbative expressions of $\mathcal{A}_{\sigma\sigma'}$ \eqqrefs{Aupdw}{Aupup} to those of
$f_{\sigma\sigma'}$ and $g_{\sigma\sigma'}$ Eqs.~(\ref{fupdw}--\ref{gupup}), we verify
the Bethe-Salpeter equations of the forward and frontal channels 
obtained non-perturbatively in Section \ref{sec:flot}. In the Bethe-Salpeter equation on $\mathcal{A}_{\upa\upa}^{\rm fwd}$ (\eqqref{BS}),
the integral on the right-hand side is a $O(a^2)$ to leading order;  
we may then replace $f_{\upa\dwa}$ and $\mathcal{A}_{\upa\dwa}^{\rm fwd}$ there by $g$. The remaining angular integration is trivial
and yields:
\be
\frac{{\mathcal{A}}^{{\rm fwd}}_{\upa\upa}(\cos\theta)}{g}=\frac{{f}_{\upa\upa}(\cos\theta)}{g}-4\pi \bar g +O(a^3)\\
\ee 
where $\bar g=(m \pF/(2\pi)^3)g$. 
Using $\lim_{\theta\to0}J(\theta)=1$ and $4\pi\bar g=2\kF a/\pi$, one easily verifies the compatibility of this result
with \eqqref{Aupup} in the limit $\theta_{14}\to0$, $\theta_{12},\theta_{13}\to\theta$.
For $\mathcal{A}_{\upa\dwa}^{\rm fwd}$, the integral term of the Bethe-Salpeter
equation is negligible (due to the cancellation of $f_{\upa\upa}$ and $\mathcal{A}_{\upa\upa}$ to order $O(a)$).
This leaves us with
\be
\frac{{\mathcal{A}}^{{\rm fwd}}_{\upa\dwa}(\cos\theta)}{g}=\frac{{f}_{\upa\dwa}(\cos\theta)}{g}+O(a^3)%\bb{\frac{m\pF}{(2\pi)^3}}^2
\ee
which is compatible with \eqqref{Aupdw} in the limit $\theta_{14}\to0$, $\theta_{12},\theta_{13}\to\theta$. 

We now turn to the Bethe-Salpeter equation in the Bogolioubov channel. In \eqqref{BetheSalpeterpaire}
for ${\mathcal{A}}^{\rm frt}_{\upa\dwa}$, we may replace the integral term
by $2\pi\bar g^2 \int_{-1}^1\dd u\text{ln}|u|=-4\pi\bar g^2$, which gives:
\be
\frac{{\mathcal{A}}^{\rm frt}_{\upa\dwa}(\cos\theta,q_{12})}{g}=\frac{ g_{\upa\dwa}(\cos\theta,\vF q_{12})}{g}-4\pi\bar g  +O(a^3)=1+\frac{2\kF a}{\pi}\bb{\text{ln}\frac{\bar q_{12}}{4}+1+J(\theta)} \label{BSBogocontact}
\ee
Note that the evaluation of $g_{\upa\dwa}$ in $\Lambda=\vF q_{12}$ converted the divergence $\propto\text{ln}\bar\Lambda$ in \eqqref{gupdw} 
into a small-angle divergence $\propto\text{ln}\bar q_{12}$ (we recall that $\bar q_{12}=2\cos(\theta_{12}/2)$). 
Expression \eqref{BSBogocontact} is compatible 
with the behavior of $\mathcal{A}_{\upa\dwa}$
when $\theta_{13}\to\theta$ and $\theta_{12}\to\pi$,
as can be seen by injecting the asymptotic behavior
$I(\theta_{12})\underset{\theta_{12}\to\pi}{=}1+\text{ln}({\bar q_{12}}/{4})$
in \eqqref{Aupdw}.
 Finally, at the working order in $g$, we may replace the integral term in the Bethe-Salpeter equation of ${\mathcal{A}}^{\rm frt}_{\upa\upa}$ by 0, such that 
\be
\frac{{\mathcal{A}}^{\rm frt}_{\upa\upa}(\cos\theta,q_{12})}{g}=\frac{ g_{\upa\upa}(\cos\theta,\vF q_{12})}{g}  +O(a^3)
\ee
The absence of a logarithmic behavior with $q_{12}$ is here an artefact of the perturbative expansion.

\subsection{The GMB correction to $T_c$ and $\Delta$}
\label{sec:GMB}

We apply here the calculation of $T_c$ from Section \ref{sec:BCS}
to the contact Fermi gas. We show that computing the $s$-wave pair interaction
$G_{\upa\dwa}^0$ to second order in $\kF a$ before applying \eqqref{Tcexacte} yields the Gor'kov-Melik Barkhudarov (GMB)
expression of $T_c$.

Let us first recall that BCS theory describes pairing of particles
under the effect of the bare interactions, which provides a first approximation of the critical temperature:
\begin{equation}
    \frac{T_{\rm c}^{\rm BCS}}{\TF} = \frac{8 \textrm{e}^{\gamma-2}}{\pi} \ \eee^{{\pi}/{2\kF a}}
\end{equation}
This corresponds to a $s$-wave pairing strength 
$1/4\pi G_{\upa\dwa}^{\rm BCS}= {\pi}/{2\kF a}-2+3\text{ln}2$ in \eqqref{Tcexacte}. However,
this is not a systematic perturbative expansion of $G_{\upa\dwa}^0$, and
only the leading term ${\pi}/{2\kF a}$ is reliable. BCS theory thus makes an uncontrolled error on the ratio $T_c/\TF$. 
To correct this BCS approximation, GMB \cite{GMB} performed a second-order diagrammatic
calculation in which they introduce in particular a dressed Green's function
and an effective interaction. 

The GMB correction is often understood \cite{Viverit2000,Chen2016,Strinati2018} as the result
of the screening of the pairing interactions among particles.
Our low-energy effective theory provides a more general interpretation of these corrections 
as the result of the renormalisation of the particles into Landau quasiparticles.
In this picture, the GMB correction arises from a systematic calculation of the effective parameters
to second order in $\kF a$.

Averaging expression \eqref{gupdw} of $g_{\upa\dwa}$ over $\theta$ yields the $s$-wave
pair interaction
\begin{equation}
    g_{\upa\dwa}^0(\Lambda) = g+g^2\frac{m^* p_{\rm F}}{2\pi^2}\bbcro{\ln{\frac{\Lambda}{\EF}}+\frac{7}{3}(1-\ln{2})}  + O(g^3)
\end{equation}
The logarithmic suppression expected from the flow equation \eqref{gfg0} is converted into a logarithmic divergence here.
We must then reexpand $1/g_{\upa\dwa}^0(\Lambda)$ in powers $g$, and identify the renormalized pairing strength $G_{\uparrow\downarrow}^0$
using \eqqref{tildeg}:
\bea
\frac{1}{4\pi G_{\uparrow\downarrow}^0} &=&  \frac{\pi}{2\kF a} + \frac{7}{3}(\ln{2}-1) +O(\kF a)\\
&=&\frac{1}{4\pi G_{\upa\dwa}^{\rm BCS}}-\frac{1}{3}\bb{1+2\text{ln}2}
\eea
This pairing strength $G_{\uparrow\downarrow}^0$ is negative and small in absolute value when $\kF a\to0^-$, which guarantees
the existence of a (weak) superfluid phase. The correction to the BCS approximation $G_{\upa\dwa}^{\rm BCS}$ is negative,
which leads to a reduction of $T_c$ and $\Delta$:
\bea
    {T^{\rm GMB}_{\rm c}}{}=&=& \frac{T^{\rm BCS}_c}{(4\eee)^{1/3}}= \frac{\textrm{e}^{\gamma}}{\pi} \left(\frac{2}{\textrm{e}}\right)^{7/3} \textrm{e}^{\pi/2\kF a} \ \TF\\
    \Delta_{\rm GMB}&=& \frac{\Delta_{\rm BCS}}{(4\eee)^{1/3}}
\eea
with $(4\eee)^{1/3}\approx 2.2$. 

Corrections to $T_c$ beyond GMB involve terms of third order in 
$1/G_{\upa\dwa}^0$ as well as the corrections to the effective mass.
This means they are not just a ``screening'' effect but result
more generally from the dressing
of the fermions into Landau quasiparticles.
These corrections are however small, \textit{i.e} of order $O(\kF a)$, in both $T_c/\TF$ and $\ln(T_c/\TF)$.

\subsection{Residue and momentum distribution}
\label{residueperturbatif}

We apply here our definition \eqqref{Zpsigma1} of the quasiparticle residue $Z_{\pp\sigma}$ to recover the perturbative expression
obtained by Mahaux et al.~\cite{Mahaux1980} (themselves correcting the incorrect result of Belyakov \cite{Belyakov1961}).

We begin with the second-order expression of the (particle) momentum distribution in an arbitrary quasiparticle state $\ket{\{n_{\pp\sigma}\}}$:
\begin{multline}
%n_{\pp\sigma}^{ \ket{\{n_{\pp'\sigma'}\}}}\equiv
\bra{\{n_{\pp'\sigma'}\}}\hat a_{\pp\sigma}^\dagger \hat a_{\pp\sigma} \ket{\{n_{\pp'\sigma'}\}}={}_0\bra{\{n_{\pp'\sigma'}\}}\hat{a}_{\pp\sigma}^\dagger \hat{a}_{\pp\sigma} + \frac{1}{2}\bbcro{[\hat{a}_{\pp\sigma}^\dagger \hat{a}_{\pp\sigma},\hat S_1],\hat S_1}\ket{\{n_{\pp'\sigma'}\}}_0+O(a^3)\\
=n_{\pp\sigma}+\bb{\frac{g}{V}}^2{\sum_{\pp_2\pp_3\pp_4\in\mathcal{D}}\delta_{\pp+\pp_2}^{\pp_3+\pp_4}{\frac{\bar n_{\pp,\sigma} \bar n_{\pp_2,-\sigma}  n_{\pp_3,-\sigma}  n_{\pp_4,\sigma}- n_{\pp,\sigma}  n_{\pp_2,-\sigma} \bar n_{\pp_3,-\sigma} \bar n_{\pp_4,\sigma}}{(\omega_{\pp}+\omega_{\pp_2}-\omega_{\pp_3}-\omega_{\pp_4})^2_{>\Lambda}}}} 
+O(a^3) \label{momentumdistrib}
%\times Q_\Lambda(\omega_{\pp}+\omega_{\pp_2}-\omega_{\pp_3}-\omega_{\pp_4}) 
\end{multline}
where 
\be
\frac{1}{(E)_{>\Lambda}^2}=\begin{cases}\frac{1}{E^2} \text{ if } E>\Lambda \\ 0\text{ else}\end{cases}
\ee 
%$Q_\Lambda(x)=1-\Pi_\Lambda(x)$. 
Applying \eqref{Zpsigma1}, we obtain
\be
Z_{\pp\sigma}
=1-\bb{\frac{g}{V}}^2{\sum_{\pp_2\pp_3\pp_4\in\mathcal{D}}\delta_{\pp+\pp_2}^{\pp_3+\pp_4}{\frac{ \bar n^0_{\pp_2,-\sigma}  n^0_{\pp_3,-\sigma}  n^0_{\pp_4,\sigma}+  n^0_{\pp_2,-\sigma} \bar n^0_{\pp_3,-\sigma} \bar n^0_{\pp_4,\sigma}}{(\omega_{\pp}+\omega_{\pp_2}-\omega_{\pp_3}-\omega_{\pp_4})^2_{>\Lambda}}}}  +O(a^3)
%Q_\Lambda(\omega_{\pp}+\omega_{\pp_2}-\omega_{\pp_3}-\omega_{\pp_4}) 
\ee
Note that the leading term in $1-Z_{\pp\sigma}$ is of order $g^2$,
in contrast to the leading interaction term in the energy $\epsilon_{\pp\sigma}$
which is of order $g$. This is because the operator $\hat{a}_{\pp\sigma}^\dagger \hat{a}_{\pp\sigma}$ (unlike $\hat H$)
is diagonal in the unperturbed basis (i.e. it commutes with the projectors $\hat P_\Lambda$).
In $p=\pF$, the residue has a well-defined $\Lambda\to0$ limit, which recovers the result of
Belyakov \cite{Belyakov1961}:
\be
    Z_{p_{\rm F}} = 1 - \frac{4(\kF a)^2}{\pi^2}\ln{2}+O(a^3)
\ee
\begin{figure}[h!]
    \centering 
    \includegraphics[width = 0.7\textwidth]{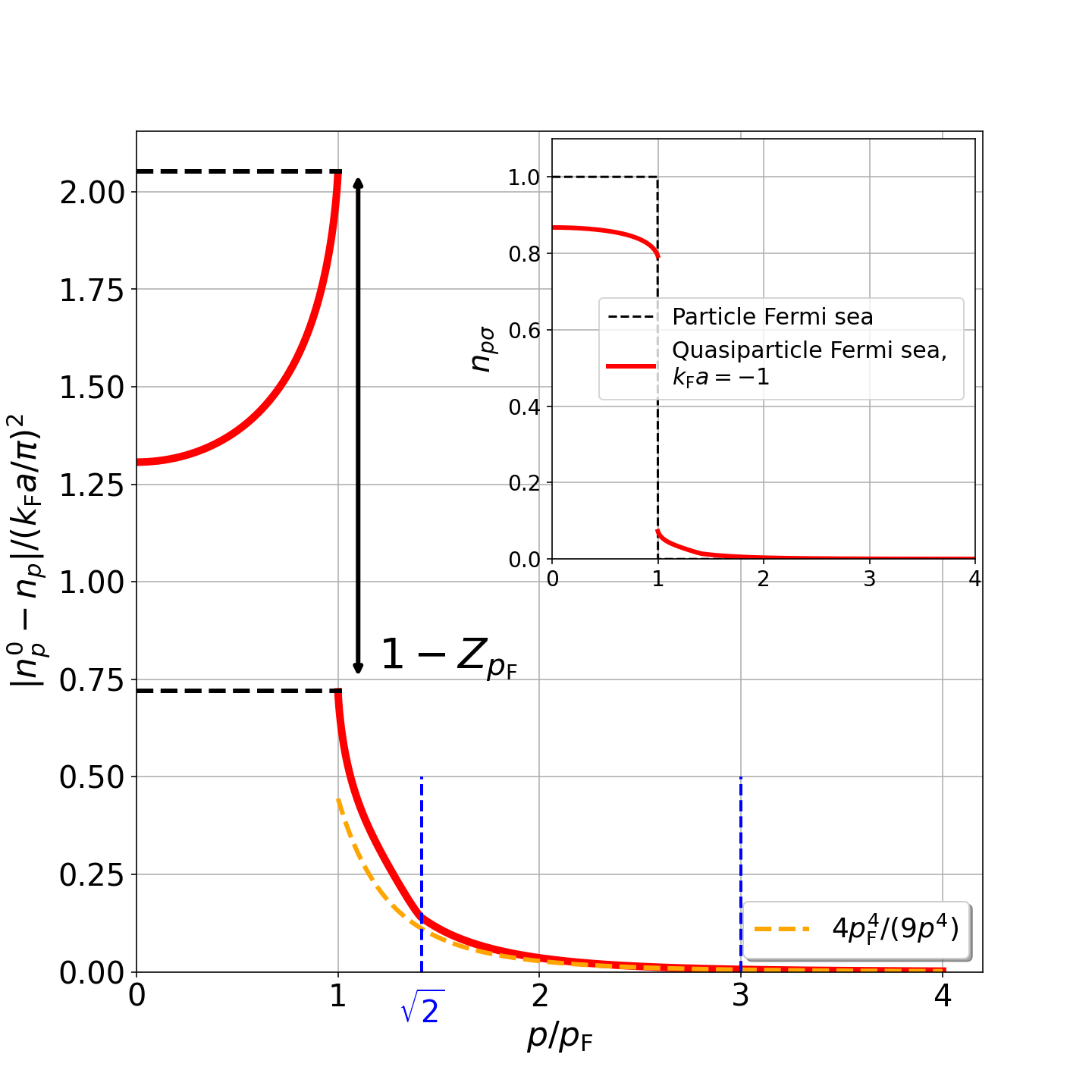}
    \caption{(Main pannel) Difference between the particle momentum distribution $n_p^{\FS}=\bra{{\rm FS}}\hat a_{\pp\sigma}^\dagger \hat a_{\pp\sigma}\FS$   
    (see \eqqref{momentumdistrib}) and the zero-temperature Fermi-Dirac distribution $n_p^0$ as a function of $p/\pF$. The difference is scaled to $(\kF a/\pi)^2$ so as to be independent of $a$ in the weak-coupling limit. 
       The discontinuity across the Fermi surface, \textit{i.e.} between the asymptotic values in $\pF-0^-$ and $\pF+0^+$ (black dashed lines), is given by 
       $1-Z_{\pF}$. At large momenta, the distribution follows a $1/p^4$ behavior, from which the contact $C = 4 (\kF a)^2 / 9\pi^2$ can 
       be extracted (orange dashed curve). (Inset) The bare distribution $n_p^{\FS}$ in function of $p/\pF$, evaluated in second-order perturbation
       theory at $\kF a=-1$. The distribution displays the familiar 
    shape of a depleted Fermi sea, with $n_{p}<1$ down to $p=0$.
 \label{fig:momentumdistribution}}
\end{figure}

Contrarily to $Z$, the momentum distribution  $n_p^{\FS}$ of the particles in the quasiparticle Fermi sea 
is well defined for all values of $p/\pF$. We depict it on Fig.~\ref{fig:momentumdistribution}
using the expressions given in \cite{Mahaux1980}, which correct the original calculation of Belyakov \cite{Belyakov1961}.
Its limiting values above and below the Fermi surface are given by
\bea
    \bra{{\rm FS}}\hat{n}_{\pF^+\sigma}\FS &=&  \frac{2(\kF a)^2}{\pi^2}\left(\ln{2}-\frac{1}{3}\right)+O(a^3) \\
    \bra{{\rm FS}}\hat{n}_{\pF^-\sigma}\FS &=&  \left[1-\frac{2(\kF a)^2}{\pi^2}\left(\ln{2}+\frac{1}{3}\right)\right] +O(a^3)
\eea
We then remark that the depletion of the particle Fermi sea is not limited to the vicinity of $\pF$
but extends all the way to $p=0$.
Then, at large momenta, the distribution decays as $1/p^4$:
\begin{equation}
     \bra{{\rm FS}}\hat{n}_{\pp\sigma}\FS  \underset{\bar p\to+\infty}{\sim} \frac{C}{\bar p^4}
\end{equation}
This provides a way to identify the first perturbative contribution to Tan's contact \cite{Tan2008}:
\be
 C = \frac{4}{9\pi^2}(\kF a)^2
\ee
Finally, besides the discontinuity in $\pF$, we note two corner points in $p/\pF=\sqrt{2}$ and 3.

\section{Zero sound in the collisionless regime}\label{sec:collless}

Having benchmarked the quasiparticle Hamiltonian with
known static properties of the contact Fermi gas, we
now turn to studying the dynamics.
Since the regime of hydrodynamic transport is covered by Ref.~\cite{devvisco}, we concentrate here on the collisionless regime $\omega_0\tau\to +\infty$, where the collision integral can be treated as a perturbation of the transport equation.
We will perform an expansion of the quasiparticle distribution 
$\nu_{\pm}$ (introduced in \eqqref{changmtvar}) in powers of $1/\omega_0 \tau$:
\be\label{devnupm}
\nu_{\pm} = \nu_{\pm}^{\rm cl}+\frac{\delta \nu_{\pm}}{\omega_0 \tau}+O(\omega_0\tau)^{-2}.
\ee
%and restrict to terms $O(1/\omega_0\tau)^0$ in the collisional correction $\delta\nu_\pm$ .

\subsection{Dispersion equation in the perfect collisionless regime ($\omega_0\tau =+ \infty$)}

Let us first compute the leading term $\nu_{\pm}^{\rm cl}$ in the perfect collisionless regime limit $1/\omega_0\tau =0$.
The low-temperature transport equation \eqref{equlowT} in this regime is
\begin{equation}
    \left(c-\cos{\theta}\right) \ \nu_{\pm}^{\rm cl}(y,\theta) = - \cos{\theta}\left(1 - \int_{-\infty}^{+\infty}\textrm{d}y'\frac{\textrm{d}\Omega'}{4\pi}F^{\pm}(\alpha)g(y')\ \nu_{\pm}^{\rm cl}(y',\theta') \right)  \label{eqintegralcolless}
\end{equation}
where $c=\omega/\omega_0$ and $\cos\alpha=\cos\theta\cos\theta'+\sin\theta\sin\theta'\cos\phi$.
Since there is no explicit dependence on the energy variable $y$ on the right-hand side, the collisionless distribution is energy-independent, $\nu^{\rm cl}_{\pm}=\nu^{\rm cl}_{\pm}(\theta)$. 
To solve the remaining 1D integral equation, we project $\nu_{\pm}^{\rm cl}$ and the interaction functions $F^{\pm}(\alpha)$ onto Legendre polynomials
\bea
\nu_{\pm}(\theta)&=&\sum_{l=0}^{+\infty}\nu^{l}_{\pm} P_l(\cos\theta) \\
F^\pm(\alpha)&=&\sum_{l=0}^{+\infty} F_{l}^{\pm} P_l(\cos\alpha) 
\eea 
which introduces the Landau parameters $F_l^\pm$. To lighten the notations, the superscript ``cl'' on $\nu$ is implicit here and until Sec.~\ref{sec:amorsonzero}. 
The integral equation folds onto a matrix equation
 whose $l$-th component is given by 
\begin{equation}\label{transportCollproj}
    \nu^{l}_\pm(c) - \sum_{l'} A_{ll'}^{\pm}(c)\nu^{l'}_{\pm}(c) + B_{l0}(c) = 0 
\end{equation}
where we have introduced the matrices
\bea
    B_{ll'}(c) &=& \int^1_{-1}\frac{\textrm{d}u}{2}P_l(u)\frac{u}{c-u}P_{l'}(u) \\
    A^{\pm}_{ll'}(c) &=& F^{\pm}_{l'}B_{ll'}(c)
\eea
This infinite-dimension linear system is  solved by formally inverting the matrix $1-A$:
\begin{equation}
    \vec{\nu}_{\pm}(c) = - \frac{1}{1-{A}^{\pm}(c)} \vec{B}_0(c)
\end{equation}
where the source vector $\vec{B}_0 = \left(B_{l0}\right)_{l \in \mathbb{N}}$ is the consequence
of the external drive (recall that $\nu$ is scaled in \eqqref{changmtvar} to the drive intensity $U$). A phononic collective
mode occurs when some component of the quasiparticle distribution $\nu$ diverges 
in response to the drive, \textit{i.e.} when the matrix $1-A^\pm(c)$ has a zero-energy
eigenvector. The dispersion equation on the reduced velocity $c_0=\omega_\qq/\vF q$ of the collective
modes is then
\begin{equation}
    \textrm{det}\left(1-{A}^{\pm}(c_0)\right) = 0,
\end{equation}
This dispersion equation can have several solutions, both real and complex. 
However, when $F_0$ is much larger than the other $F_l$'s, as in the case
of weakly-interacting Fermi gases and ${}^3$He, 
%or when the $F_{l\geq 2}$ are set to 0,
%as in most models of ${}^3$He, 
there is a dominant real solution,
traditionally called zero sound. Physically, this solution describes a longitudinal collisionless
phononic branch.

\subsection{Log-perturbative expansion of the zero-sound velocity}\label{logperturbative}
We are now calculating the zero-sound reduced frequency $c_0$ in powers of $\ab=\kF a$
in a weakly-interacting Fermi gas.
In equation \eqref{transportCollproj} for $l>0$, the summation
is dominated by the term $l'=0$ (which contains the dominant coefficient $F_0$)
so that:
\begin{equation}\label{nulpm}
    \nu^{l}_{\pm} = -B_{l0}(c) + F_{0}^{\pm}B_{l0}(c) \ \nu^{0}_{\pm} +O(\ab),\quad\text{for } l\geq 1
\end{equation}
Anticipating on the followings, we have estimated $B_{ll'}=O(1/\ab)$.
Reinjecting in \eqref{transportCollproj} for $l=0$, we eventually obtain $\bar{\chi}_{\pm}\equiv \nu^{0}_{\pm}$, which represents either the dimensionless density $\bar{\chi}_{\rho}$ or polarization $\bar{\chi}_{\rm p}$ response, depending on the $\pm$ index:
\begin{equation}\label{chipm}
    \barre{\chi}_{\pm}(c) = -\frac{B_{00}(c) +\sum_{l'>0} F^{\pm}_{l'} B_{0,l'}(c) B_{l',0}}{1-F^{\pm}_0 B_{0,0}(c)-\sum_{l'} F^{\pm}_0 F^{\pm}_{l'}B_{0,l'}(c) B_{l',0}(c)}
\end{equation}
The dispersion relation $1/\barre{\chi}(c_0) = 0 $ now reduces to:
\begin{equation}\label{dispersioncollless}
    1-F^{\pm}_0 B_{0,0}(c_0^\pm)-\sum_{l'} F^{\pm}_0 F^{\pm}_{l'}B_{0,l'}(c_0^\pm) B_{l',0}(c_0^\pm) = 0
\end{equation}
Since we expect that $c_0-1 $ tends to zero exponentially as $|\ab| \rightarrow 0$, we introduce the variable $\gamma^\pm = \ln{[(c_0^\pm-1)/2]}$. Contrarily to $c_0$, $\gamma$ can be expanded in power of $\ab$: 
\begin{equation}
    \gamma^{\pm} = \frac{\gamma_0^{\pm}}{\ab} + \gamma_1^{\pm} + {O}(\ab)
\end{equation}
The log-perturbative corrections to $\gamma$ convert into a prefactor correction in $c_0-1$,
reminiscent of the Gor'kov Melik-Barkhudarov prefactor in the calculation of the superfluid critical temperature \cite{GMB}.
When $c_0$ tends to 1 exponentially, the functions $B_{ll'}$ have the following expansion 
\begin{equation}
    B_{00}(c_0^\pm) = -1 -\frac{1}{2}\gamma^{\pm} + O(\ab) \ \textrm{ and } \ B_{l0}(c_0^\pm)=B_{0l}(c_0^\pm) = \frac{\gamma^{\pm}}{2} +O(1)
\end{equation}

By substituting the expansions of $\gamma$, of the Landau parameters $F^{\pm}_{l}$ and of the functions $B_{ll'}$ 
into \eqref{dispersioncollless}, and by restricting to terms of order $O(\ab)$, we obtain the following expressions of $\gamma_0^{\pm}$ and $\gamma_1^{\pm}$:
\begin{equation}\label{gammadispersioncoll}
    \gamma_0^{\pm} = \mp\pi  \ \textrm{ and } \ \gamma_1^{\pm} = -2 +\frac{\pi^2}{2\ab^2}\left(\sum_{l>0} F^{\pm}_{l} +\delta F_0^{\pm}\right) =\pm 4 
\end{equation}
where we have expanded $F_0^{\pm}$ as:
\begin{equation}
    F_0^{\pm} = \pm\frac{2\ab}{\pi} + \delta F_0^{\pm}, \quad \delta F_0^{\pm}=O(\bar a^2) 
\end{equation}
We eventually recognize in $\gamma^{\pm}$ the sum of the $F_l^{\pm}$ that is the forward value ($\alpha=0$) of $F^{\pm}$.
\begin{equation}
    \gamma^{\pm} = -\frac{2}{F^{\pm}(\alpha=0)} -2 +O(\ab)
\end{equation}
This shows that the Landau function $F(\alpha)$ in the integral equation \eqref{eqintegralcolless} 
can be replaced (to leading and subleading order in $\ab$) by its value in $\alpha=0$, that is for quasiparticles with
collinear momenta $\pp\parallel\pp'$.
This is a consequence of the longitudinal nature of zero sound at weak-coupling:
the quasiparticle distribution $\nu_\pm(\theta)\propto \cos\theta/(c_0-\cos\theta)$
is peaked about $\theta=0$, such that the quasiparticle momenta $\grp$ are all nearly
collinear to $\gq$.

In short, the zero sound velocity (in units of $\vF$) for the density mode is given by 
\begin{equation} \label{c0p}
    c_0^+ = 1 + 2 \textrm{e}^{4} \ \textrm{e}^{-\pi / \ab}, \qquad \ab >0
\end{equation}
and the velocity for the zero polarization mode is:
\begin{equation} \label{c0m}
    c_0^- = 1 +2 \textrm{e}^{-4} \ \textrm{e}^{\pi / \ab}, \qquad \ab <0
\end{equation}
We recall that the leading-order result \cite{FetterWalecka,yaleexp}, corresponding to the Random Phase Approximation (RPA), 
is $c_{0,{\rm RPA}}^\pm=1+2\eee^{-2}\textrm{e}^{\mp\pi / \ab}$,
which amounts to setting $\gamma_1^{\pm} = -2$ in \eqqref{gammadispersioncoll}.
Second order corrections thus shift the density zero sound to higher velocities in the density-density response, increasing $c_0^+-1$ by a factor $\textrm{exp}(6)\simeq 403$.
We then expect the resonance to be more easily observable than predicted by first-order approximations.
Since it exists only for $\ab>0$, the density zero sound is observable in a Fermi gas only on the metastable branch. Experimental exploration of this metastable branch are restricted to $|\ab|\lesssim0.1$, where zero
sound is visible only at very low temperatures.

Conversely, the polarisation sound mode, which is observable on the ground branch at $\ab<0$, is shifted closer to the continuum edge, with $c_0^- -1$ 
reduced by a factor $\textrm{exp}(-2)\simeq0.14$. This reduces the temperature range in which this zero sound mode is observable. 

We have benchmarked these analytical results using a numerical solution of \eqqref{transportCollproj}.
More details for the numerical evaluation are given in Appendix \ref{app:sonzero}.

\subsection{Response function in the collisionless regime}
Our discussion of zero sound so far has focused on reduced frequencies $c\approx c_0\approx 1$. We now discuss numerically the rest of the spectrum in the density-density response $\textrm{Im}[\barre{\chi}_{\rho}] = \textrm{Im}[\nu^0_{+}]$ and polarisation-polarisation response $\textrm{Im}[\barre{\chi}_{p}] = \textrm{Im}[\nu^0_{-}]$. 
 Figs.~\ref{Lindarddensity} and \ref{Lindardpolarisation} show the reduced spectral density $\textrm{Im}[\barre{\chi}_{\rho,\rm p}(c+i0^+)]$.
In the attractive case $\ab<0$ (red curves), second-order corrections tend to decrease the deviations of $\textrm{Im}[\barre{\chi}_{\rho}]$ and $\textrm{Im}[\barre{\chi}_{\rm p}]$  from the Lindhard response of an ideal gas (black curve). 
Since a stable Fermi liquid regime exists in ultracold Fermi gases only for $\ab < 0$, this behavior should be the easiest to observe in cold atom experiments.
Conversely, in the repulsive case $\ab>0$ (blue curves), the deviations are increased to second-order.
This can be understood by comparing the first- and second-order approximation of the Landau parameters. For the leading coefficient, $F^\pm_0$, we have:
\bea
    \frac{F^{+(2)}_0}{F^{+(1)}_0} &\simeq& 1+ 1.143 \ \ab \\
        \frac{F^{-(2)}_0}{F^{-(1)}_0} &\simeq& 1 + 0.130 \ \ \ab
\eea
Thus, for negative (resp. positive) $\ab$, the second-order $F_0^\pm$ is smaller (resp. larger) than its first-order counterpart. 
 As a result, the effective interaction between quasiparticles is reduced (resp. increase), tending to restore (resp. remove) the behavior of an ideal gas. 
 Even though this is true both in the
density and polarisation channel, the effect is $\approx10$ times larger in the density channel.

In the density response (Fig.~\ref{Lindarddensity}), a zero sound resonance appears, in the repulsive case, as a Dirac peak at $c_0>1$; there remains also a secondary peak near the edge of the quasiparticle-quasihole continuum for $c\lesssim 1$ (see inset).
This secondary peak visibly shrinks as second-order corrections push the zero sound resonance away from the continuum. In the attractive case, interactions tend to smoothen the sharp behavior at the continuum edge, and the density response
becomes a broad, featureless spectral function.

\begin{figure}[htb]
    \centering
    \includegraphics[width=0.75\linewidth]{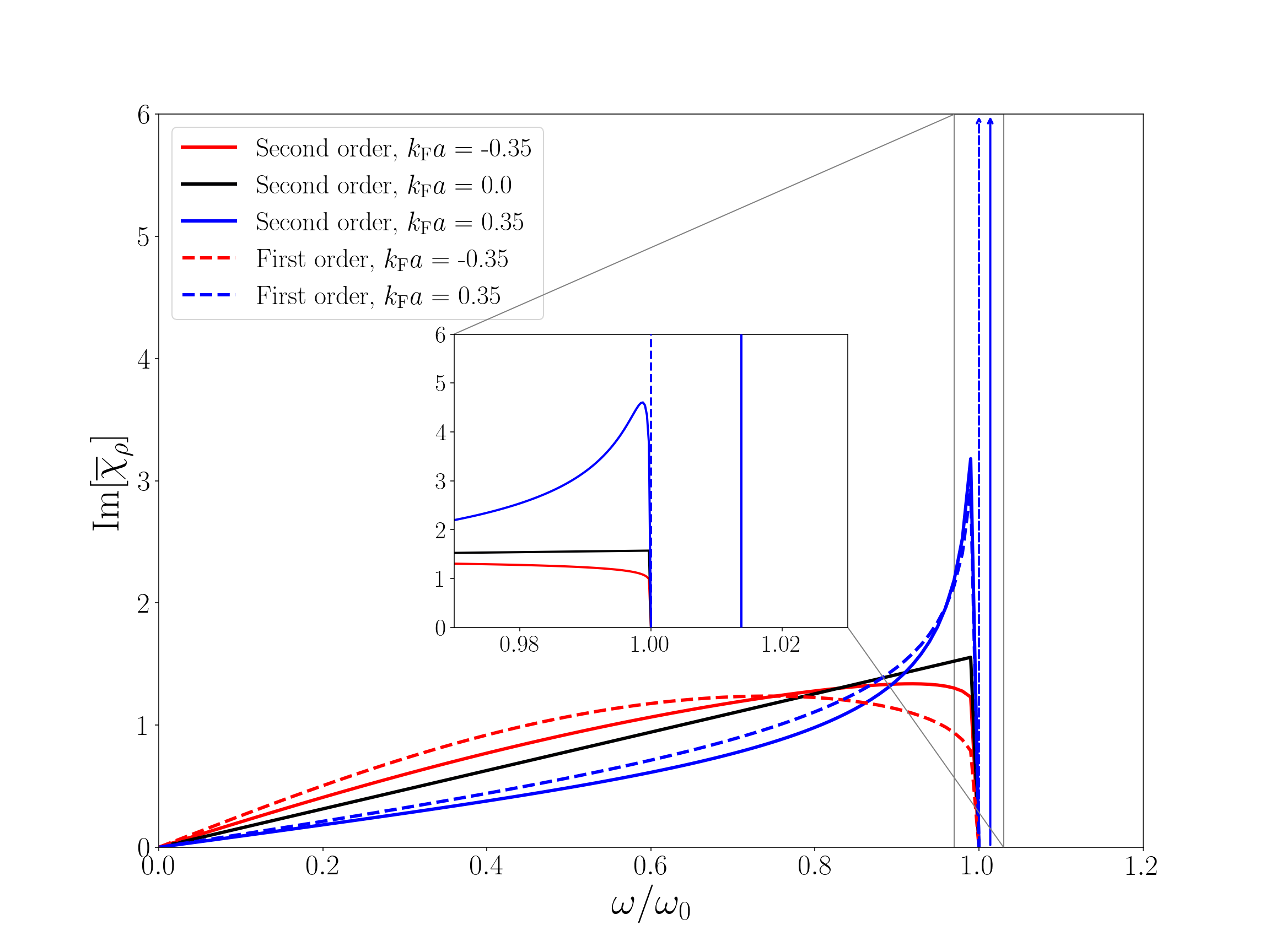} 
    \caption{The reduced spectral density $\textrm{Im}[\barre{\chi}_{\rho}(c+\ii0^+)] = \textrm{Im}[\nu_{0}^{+}(c+\ii0^+)]$ as a function of $c=\omega/\omega_0$, for different values of $\ab=k_{\rm F}a$, blue curves for $\ab>0$ and red curves for $\ab<0$. The dashed lines correspond to the first order calculation \cite{theserepplinger}. The black line is the non-interacting case. The solid curves include second-order effects; they are obtained by numerically solving \eqref{transportCollproj} truncated to $l_{\rm max}= 100 $.}
    \label{Lindarddensity}
\end{figure}

Conversely, in the polarisation response (Fig.~\ref{Lindardpolarisation}), the resonance appears in the attractive case, and the broad structure in the repulsive case,
which indicates a repulsive/attractive, density/polarisation duality.
This time, the resonance is brought closer to the continuum edge by second-order corrections, and the secondary peak near the continuum edge grows.
In presence of a small spectral broadening (either due to collisional damping, Landau damping or experimental resolution) the two peaks would become indistinguishable.
%\\

\begin{figure}[htb]
    \centering
    \includegraphics[width=0.75\linewidth]{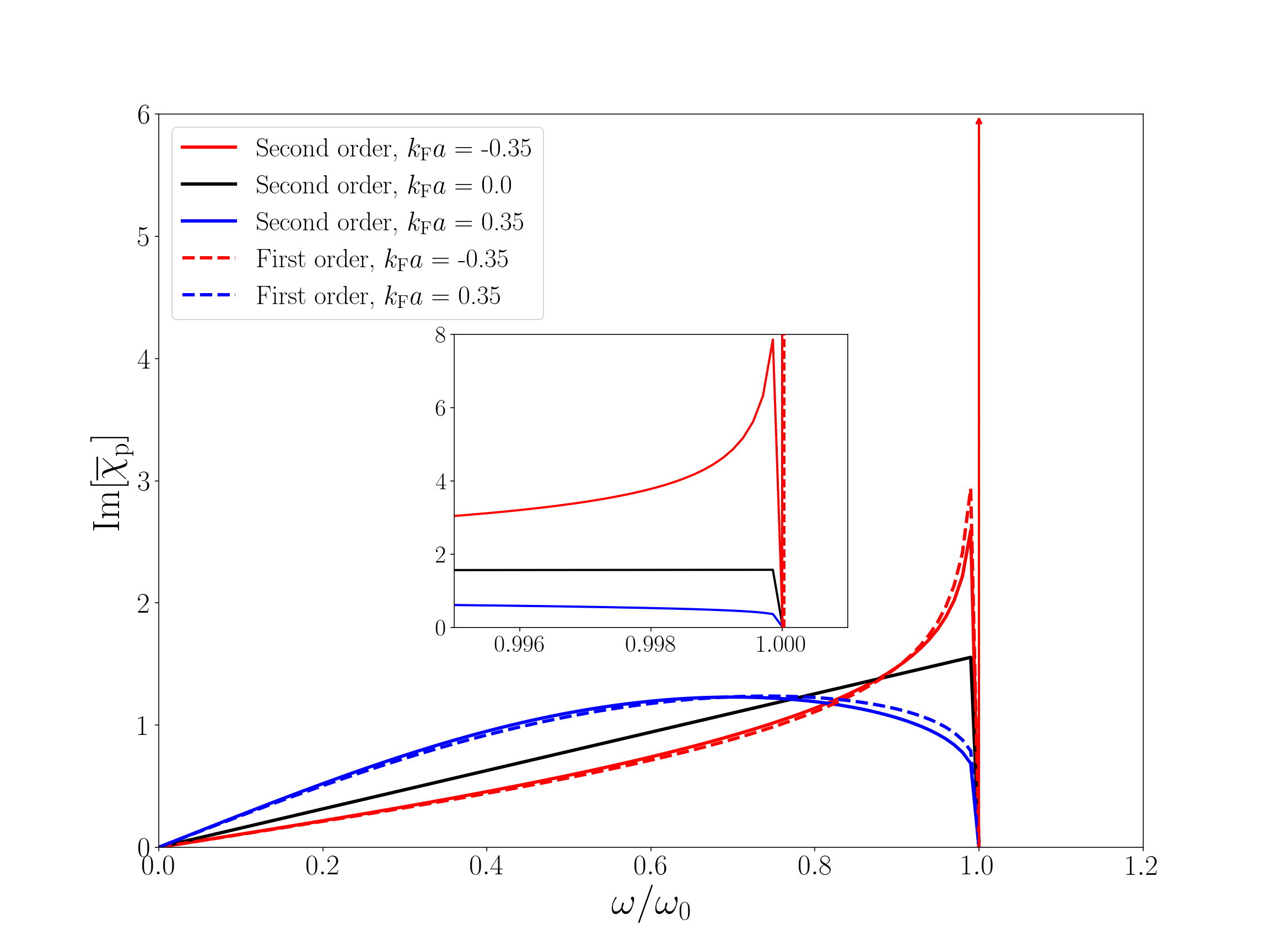} 
    \caption{The reduced spectral density for the polarisation $\textrm{Im}[\barre{\chi}_{\rm p}(c+\ii0^+)] = \textrm{Im}[\nu^{-}_{0}(c+\ii0^+)]$ for the same parameters as Fig.~\ref{Lindarddensity}.}
    \label{Lindardpolarisation}
\end{figure}

\subsection{Collisional damping of zero sound} \label{sec:amorsonzero}
We now aim to include the collisional correction $\delta\nu_\pm$ (see \eqqref{devnupm}) in the distribution $\nu_{\pm}$.
In the regime where $c$ is exponentially close to 1 ($\gamma=\ln([c-1]/2)\to -\infty$), the leading-order solution $\nu_{\pm}^{\rm cl}$ is more easily written without the Legendre decomposition, directly in terms of the angular variable $\theta$:
\begin{equation}
    \nu^{\rm cl}_{\pm}(\theta) = \frac{\cos{ \theta}}{c-\cos{\theta}} \cdot\frac{\rho^{{\rm cl}}_{\pm}(c)}{B_{00}(c)},\qquad \rho_{\pm}^{\rm cl}(c)= \frac{- B_{00}(c)}{1-F^{\pm}(0)B_{00}(c)}
\end{equation}
This solution was obtained by replacing $F^{\pm}(\alpha)$ with $F^{\pm}(0)$ in \eqqref{eqintegralcolless}, as shown in Section \ref{logperturbative}.
To simplify the notation, we denote by $\rho^{\rm cl}_{\pm}$ the $l=0$ component of $\nu_{\pm}^{\rm cl}$, corresponding
to density/polarisation fluctuations:
\begin{equation}
     \rho^{\rm cl}_{\pm} = \nu_\pm^{0,{\rm cl}}= \int_0^\pi \frac{\sin\theta\textrm{d}\theta}{2} \ \nu_{\pm}^{\rm cl}(\theta)
\end{equation}
We define $\rho_{\pm}=\rho_{\pm}^{\rm cl}+\delta \rho_{\pm}^{\rm cl}$ in the same way.
We then substitute the expansion of $\nu_{\pm}$ given in \eqqref{devnupm} into the transport equation \eqref{equlowT}, keeping terms up to order $1/\omega_0\tau$.
Since $\nu_{\pm}^{\rm cl}$ does not depend on energy, we average out the energy dependence of the collision integral:
\begin{equation}
    \int^{+\infty}_{-\infty} \textrm{d}y' \ \mathcal{S}(y,\pm y') = \barre{\Gamma}(y), \qquad  \int^{+\infty}_{-\infty} \textrm{d}y \ g(y)\bar\Gamma(y)=\frac{4\pi^2}{3}
\end{equation}
We then obtain the following equation for $\nu_{\pm}$:
\begin{equation}
    \nu_{\pm}(\theta) = -\frac{\cos{\theta}}{c-\cos{\theta}}\left(1-{F^{\pm}(0)} \rho_{\pm}\right) - \frac{4\pi^2}{3}\frac{\ii}{\omega_0 \tau}\frac{1}{c-\cos{\theta}}\left(\nu^{\rm cl}_{\pm}(\theta) + \int \frac{\textrm{d}\Omega'}{2\pi}\, \mathcal{N}_{\pm}(\alpha) \, \nu^{\rm cl}_{\pm}(\theta')\right) 
\end{equation}
where the angular collision kernel $\mathcal{N}_{\pm}(\alpha)$ and its expansion in Legendre polynomials are given in Appendix \ref{app:colldamp}.
In the collision integral, we have replaced the deviation from local equilibrium $\Phi$ by the deviation to true equilibrium $\nu$. This is justified in this weakly-interacting limit, where one can approximate (as in the previous subsection) $F^{\pm}(\alpha)$ by $F^{\pm}(0)$ in the relation between $\Phi$ and $\nu$ \eqqref{phinu}. Thus, $\Phi$ differs from $\nu$ only by a constant:
\begin{equation}
    \Phi_{\pm}^{\rm cl}(\theta) - \nu_{\pm}^{\rm cl}(\theta) = F^{\pm}(0)\rho_{\pm}^{\rm cl}
\end{equation}
which belongs to the kernel of the collision integral.

We then integrate over $\theta$ so as to obtain $\rho_{\pm}$ on the left-hand side.
In the collision integral, $\rho_{\pm}^{\rm cl}$ can be replaced by the total density $\rho_{\pm}$, neglecting terms of order $1/(\omega_0\tau)^2$.
We obtain the following solution:
\begin{equation}
     \rho_{\pm}(c)= -\frac{ B_{00}(c)}{1-F^{\pm}(0)B_{00}(c)+\frac{\ii}{\omega_0 \tau}\frac{2\pi^2}{3}\frac{C_{\pm}(c)}{B_{00}(c)}}
\end{equation}
where the collisional contribution $C_{\pm}$ is given by:
\begin{equation}\label{CollisionCollessKernel}
    C_{\pm}(c) = \int^{\pi}_0 \textrm{d}\theta \ \frac{\sin{\theta}\cos{\theta}}{(c-\cos{\theta})^2} + \int \frac{\textrm{d}\Omega'}{2\pi} \textrm{d}\theta \ \frac{\sin{\theta}\cos{\theta'}}{(c-\cos{\theta})(c-\cos{\theta'})}  \mathcal{N}_{\pm}(\alpha) 
\end{equation}
Since these integrals are dominated by the vicinity of $\theta = 0$ and $\theta' = 0$, the collision kernel $\mathcal{N}_{\pm}(\alpha)$ can be replaced by its value at $\alpha = 0$.
This can again be interpreted as a consequence of the quasi-longitudinal nature of zero sound in the weak-interaction regime.
We thus arrive at
\begin{equation}
    C_{\pm}(c) \simeq -B_{00}'(c) \left(1+2(c-1)\gamma^2(c) \ \mathcal{N}_{\pm}(0)\right)
\end{equation}
with $\gamma(c) = \ln{\frac{c - 1}{2}}$ and $B_{00}'(c)=\dd B_{00}/\dd c$.
A more general calculation of the function $C_{\pm}$ is given in Appendix \ref{app:colldamp}.

%A l'ordre dominant, $C_{\pm}(c) \simeq -B_{00}'(c)$, et acquiert donc un comportement universel pour tout liquide de Fermion en interaction faibles.

To obtain the collisional correction to the zero-sound velocity, we now solve the equation
\begin{equation}
\frac{1}{ \rho_{\pm}(z_0^\pm)} = 0, \quad \text{with } z_0^\pm = c_0^\pm + {\delta c_{0}^{\pm}}.
\end{equation}
Expanding the denominator of $\rho_{\pm}$ in powers of $O(1/\omega_0\tau)$, we finally extract the collisional correction to zero sound, which is purely imaginary:
\begin{equation}
\delta c_{0}^{\pm} = \frac{2\pi^2}{3} \frac{\ii}{\omega_0\tau}\frac{C_{\pm}(c_{0}^{\pm})}{B'_{00}(c_0^{\pm})} = -\frac{2\pi^2}{3} \frac{\ii}{\omega_0\tau} +O(c_0-1).
\end{equation}
where we have used $C_{\pm}(c_0^\pm)\simeq -B_{00}'(c_0^\pm)+O(c_0-1)$ for the last equality.

This result describes the broadening of the zero-sound resonance in the response functions $\chi_\rho(c)$ and $\chi_p(c)$,
or equivalently, its exponential damping in the time domain.
To leading order in $c_0-1$, $\mathrm{Im}(c_0)$ depends on the collision probability $W$ only through the mean collision time $\tau$,
which makes the product $\omega_0\tau\times \mathrm{Im}(c_0)$ universal in weakly-interacting Fermi liquids.
In this sense, the damping of zero sound differs from that of hydrodynamic sound (the first sound),
which is sensitive —via the shear viscosity $\eta$— to the angular dependence of $W$,
and therefore varies with $\kF a$ in a way that differs significantly from $\tau$.

%Dans la limite où $c$ tends vers $1$ exponentiellement, comme dans un liquide de fermions en interaction faible, $\delta c_{\pm}$ acquiert une valeur universelle:
%\begin{equation}
%    \delta c_{\pm}^{\rm univ} = - \textrm{i} \frac{2\pi^2}{3}
%\end{equation}

\section{Numerical solution in the collisionless to hydrodynamic crossover}\label{sec:collless2}
Between the collisionless regime studied in Section \ref{sec:collless} and the hydrodynamic regime treated in Ref.~\cite{devvisco}, there exists a smooth transition as a function of $\omega_0\tau$ \cite{Khalatnikov1958,Pethick1977}.
In the following, we develop a numerical method that solves the transport equation \eqref{equlowT} in this intermediate regime.
This method is based on the orthogonal decomposition \cite{Schafer2010,Hofmann2023,Hofmann2025} of the quasiparticle distribution $\nu$ developed in Refs.~\cite{disphydro,devvisco},
that we recall below. This decomposition is numerically much more efficient than a mere discretisation on a momentum grid,
and it allows for a clear control of the numerical errors.

\subsection{Decomposition in an orthogonal basis}
\label{decomposition}

%While the Legendre basis for the angular dependence is
%evident, we construct a specific family $Q_n$ for the energy dependence.
%Orthogonal decompositions are frequently introduced 
%to perform exact inversions of the collision kernel, as we shall do in the next chapter.

\subsubsection{Angular decomposition over Legendre polynomials}
Let us recall 
the angular decomposition of $\nu$ over the Legendre polynomials $P_l$
\begin{equation} \label{decomponul}
\nu_\pm(y,\theta)= \sum_{l\in\mathbb{N}} \nu_{\pm}^l(y) P_l(\cos\theta) 
\end{equation}
The spherical addition theorem ensures that angular momentum $l$ is conserved during
the collision, in other words, that the Legendre basis diagonalizes the angular part of the collision kernel.
Let us illustrate this by projecting \eqqref{equlowT} on $P_l$:
%\begin{multline}
%{\frac{\omega}{\omega_0}}{\nu}_\pm^l(y)-\frac{l}{2l-1}\bb{{\nu}_\pm^{l-1}(y)+ \frac{F_{l-1}^\pm}{2l-1}\nu_{0\pm}^{l-1}}-\frac{l+1}{2l+3}\bb{{\nu}_\pm^{l+1}(y)+\frac{ F_{l+1}^\pm}{2l+3}\nu_{0\pm}^{l+1}}\\
%+\frac{\ii}{\omega_0\tau}\bbcro{\bar \Gamma(y)\Phi^l_\pm(y)+\frac{2}{2l+1}\int_{-\infty}^{+\infty}\dd y' \bb{\mathcal{S}(y,-y')\frac{\Omega_{E\pm}^l}{\Omega_\Gamma}-2\mathcal{S}(y,y')\frac{\Omega_{S\pm}^l}{\Omega_\Gamma}}\Phi_{\pm}^l(y')}=-\delta_{l1}
%\label{nul}
%\end{multline}
\begin{multline}
{\frac{\omega}{\omega_0}}{\nu}_\pm^l(y)-\frac{l}{2l-1}{{\Phi}_\pm^{l-1}(y)}-\frac{l+1}{2l+3}{{\Phi}_\pm^{l+1}(y)}\\
+\frac{\ii}{\omega_0\tau}\bbcro{\bar \Gamma(y)\Phi^l_\pm(y)+\frac{2}{2l+1}\int_{-\infty}^{+\infty}\dd y' \bb{\mathcal{S}(y,-y')\frac{\Omega_{E\pm}^l}{\Omega_\Gamma}-2\mathcal{S}(y,y')\frac{\Omega_{S\pm}^l}{\Omega_\Gamma}}\Phi_{\pm}^l(y')}=-\delta_{l1}
\label{nul}
\end{multline}
where ${\Phi}_\pm^{l}(y)={\nu}_\pm^{l}(y)+ \frac{F_{l}^\pm}{2l+1}\int_{-\infty}^{+\infty}\dd y' g(y')\nu_{\pm}^{l}(y')$.
%$\nu_0^{l\pm 1}$ is the energy average of $\nu^{ l\pm 1}(y)$ \eqref{decomp_nunl}.
The function $\nu^l$ is coupled through collisions (second line of \eqqref{nul}) only to itself. It is however coupled to $\nu^{l+1}$ and $\nu^{l-1}$
through the streaming terms of the Boltzmann equation, which organizes the spherical harmonics in a ladder.

%We have introduced the Landau parameters
%\begin{equation} \label{Fl}
%F_l^\pm=\frac{m^*p_{\rm F}}{\pi^2}\int_0^\pi\sin\theta\dd\theta\frac{P_l(\cos\theta)}{||P_l||^2}\frac{f_{\upa\upa}(\cos\theta)\pm f_{\upa\dwa}(\cos\theta)}{2}
%\end{equation}
%\textit{i.e.} the decomposition of $f_{\upa\upa}\pm f_{\upa\dwa}$ onto the Legendre polynomials $P_l$. 
%We will use an alternative
%notation
%\be \label{barF}
%\tilde F_l^\pm\equiv1+\frac{F_l^\pm}{2l+1}
%\ee
%for this combination that frequently appears in the transport equation. 
%In these notations, the effective mass is simply
%\be
%\frac{m^\ast}{m}=\tilde F_1^+
%\ee
Similarly to the decomposition of $f$, we have decomposed the (azimuthally-integrated) collision probabilites $\Omega_{E\pm}$ and $\Omega_{S\pm}$ \eqqrefs{OmegaEalpha}{OmegaSalpha}:
\begin{eqnarray}
\frac{\Omega_{E\pm}^l}{2l+1}&=&\meanvlr{\frac{W_{E\pm}(\theta,\phi)P_l(\cos\theta)}{2\cos(\theta/2)}}_{\theta,\phi} \label{OmegaE}\\
\frac{\Omega_{S\pm}^l}{2l+1}&=&\meanvlr{\frac{W_{S\pm}(\theta,\phi)P_l(1-2\sin^2\frac{\theta}{2}\sin^2\frac{\phi}{2})}{2\cos(\theta/2)}}_{\theta,\phi}\label{OmegaS} 
\end{eqnarray}
with the solid-angular average $\meanv{f}_{\theta,\phi}=\int_0^\pi \int_0^{2\pi} f(\theta,\phi)\sin\theta\dd\theta\dd\phi/4\pi $. 
%In the isotropic case ($W_{\upa\dwa}=\text{cte}$) one has $\Omega_{S+}^l=\Omega_{E}^0$, $\Omega_{S-}^l=0$ and $\Omega_{E}^l=(-1)^l \Omega_{E}^0$.
%: ${\tau}^{-1}={(m^*)^3T^2}\Omega_E^0/{(2\pi)^3}$.

\subsubsection{Scalar product and orthogonal polynomials for the energy dependence}
To decompose the remaining dependence on $y$, 
we construct a family of polynomials $\{Q_n\}_{n\in\mathbb{N}}$
orthogonal for the scalar product:
\be
\int_{-\infty}^{+\infty} g(y) Q_n(y) Q_m(y)\dd y =||Q_n||^2\delta_{nm}.
\ee
The polynomials $Q_n$ are obtained by the usual recurrence relation:
\be
y Q_n=Q_{n+1}+\xi_n Q_{n-1} \quad\text{with}\quad\xi_n\equiv\frac{||Q_n||^2}{||Q_{n-1}||^2}
\ee
We choose $Q_0=1$ and $Q_1=y$ as the initial condition.
Note that even and odd polynomials are respectively symmetric
and antisymmetric about the Fermi surface: $Q_n(-y)=(-1)^nQ_n(y)$.
The full decomposition of $\nu$ reads
\begin{equation} \label{decomp_nunl}
\nu_\pm(y,\theta)= \sum_{l,n\in\mathbb{N}} \nu_{n\pm}^l P_l(\cos\theta) Q_n(y),\qquad \nu_{n\pm}^l=\int_{0}^{\pi}\sin\theta\dd\theta\int_{-\infty}^{+\infty} \dd y\frac{P_l(\cos\theta) Q_n(y)}{||P_l||^2 ||Q_n||^2}\, \nu_\pm(y,\theta) 
\end{equation}
The weighting of the scalar product by $g(y)=-T(\dd n_{\rm eq}/\dd \epsilon)$
makes sure that the conserved quantities \eqqref{deltarho}--\eqqref{deltae} coincide with single 
components of $\nu$:
\bea
\delta\rho_\pm&=&-\bb{U_\pm \frac{m^\ast\pF}{2\pi^2}} \nu_{0\pm}^0 \label{consquant1}\\
 m v_\parallel&=&-\bb{ U_+\frac{m^\ast}{2\pF}}\nu_{0+}^1\label{consquant2} \\
\delta e&=&-\bb{U_+\frac{m^\ast T\pF}{6}}\nu_{1+}^0 \label{consquant3}
\eea
where $\delta\rho_\pm=\delta\rho_\upa\pm\delta\rho_\dwa$.
This makes our orthogonal basis particularly well suited to the hydrodynamic regime.

Let us decompose the energy functions $\bar \Gamma$ and $\mathcal{S}$ of the collision kernel over the $Q_n$:
\bea
\Gamma_{nn'}&=&\int_{-\infty}^{+\infty} \dd y g(y)\bar \Gamma(y) \frac{Q_n(y)}{||Q_n||^2} Q_{n'}(y) \label{Gammannp_int} \\
&=&\bb{\pi^2+\xi_{n+1}+\xi_{n}}\delta_{nn'}+\delta_{n-2,n'}+{\delta_{n+2,n'}}\xi_{n+2}\xi_{n+1} \label{Gammannp}\\
\mathcal{S}_{nn'}&=&\int_{-\infty}^{+\infty} \dd y \dd y' g(y)\mathcal{S}(y,y') \frac{Q_n(y)}{||Q_n||^2} Q_{n'}(y') \\
&=&{2\pi^2\frac{n^2+n-1}{4n^2+4n-3} }\delta_{nn'}+\frac{\delta_{n-2,n'}}{n(n-1)}+\frac{\delta_{n+2,n'}\xi_{n+2}\xi_{n+1}}{(n+2)(n+1)} \label{Snnp}
\eea
with $\xi_n=\frac{\pi^2 n^4}{(2 n + 1)(2 n - 1)}$ as can be shown recursively.
Note that the subspaces of symmetric and antisymmetric  functions of $y$ (even and odd $n$ respectively)
are decoupled: $\Gamma_{nn'}=\mathcal{S}_{nn'}=0$ if ${n+n'}$ is odd \cite{Sykes1970}.
$\Gamma$ and $\mathcal{S}$ are thus represented by a tridiagonal matrix in
each of the even and odd sector.

The final decomposition of the transport equation reads
\be
\bb{\frac{\omega}{\omega_0}+\frac{\ii}{\omega_0\tau}\mathcal{M}_\pm^l \mathcal{F}^{l}_{\pm}}\vec{\nu}_\pm^l-\frac{l}{2l-1} \mathcal{F}^{\pm}_{l-1}\vec{\nu}_\pm^{l-1}-\frac{l+1}{2l+3} \mathcal{F}^{\pm}_{l+1} \vec{\nu}_\pm^{l+1}=-\delta_{l1}\vec{u}_0
\label{eq:transportprojBT}
\ee
where we treat the $n$-dependence using vector/matrix notations
\begin{equation}
    \vec{\nu}^l_{\pm} = 
    \begin{pmatrix}
        \nu_{0\pm}^l \\
        \nu_{1\pm}^l \\
        \vdots
    \end{pmatrix},
    \qquad  \vec{u}_0 =
    \begin{pmatrix}
        1 \\
        0\\
    \vdots
    \end{pmatrix},
    \qquad \mathcal{U}_0 =
    \begin{pmatrix}
        1 & 0 & 0 & \cdots \\
        0 & 0 & 0 & \cdots  \\
        \vdots 
    \end{pmatrix}
\end{equation}
\begin{equation}
    \mathcal{F}^{\pm}_l = \left(1+\frac{F^{\pm}_l}{2l+1}\mathcal{U}_0\right) =
    \begin{pmatrix}
        1 + \frac{F_l^{\pm}}{2l+1} & 0 & 0 & \cdots \\
        0 & 1 & 0 &  \cdots \\
        \vdots & & \ddots \\
        0 & \cdots & \cdots & 1
    \end{pmatrix}, \qquad    \vec{\Phi}^l_{\pm} =  \mathcal{F}^{\pm}_l \vec{\nu}^l_{\pm}
\end{equation}
% $\vec{\nu}_\pm^l=(\nu_{n\pm}^l)_{n\in\mathbb{N}}$, $\vec{u}_0=(\delta_{n0})_{n\in\mathbb{N}}$ and $\mathcal{U}_0\vec\nu=(\vec{u}_0\cdot\vec\nu)\vec{u}_0$.
%One can show that we have the following relation between $\vec{\Phi}$ and $\vec{\nu}$ when applied on the collision matrix $\mathcal{M}^l$:

The matrix $\mathcal{M}^l$ contains the contribution of the collision integral to the transport equation of the $l$th harmonic; it is a linear combination of the matrices
$\Gamma$ and $\mathcal{S}$
\begin{equation}
\mathcal{M}_{nn'\pm}^l\!=\!
\begin{cases}
\mathcal{M}_{nn'}(\alpha_{n\pm}^l)\equiv \Gamma_{nn'}-\alpha_{n\pm}^l \mathcal{S}_{nn'} \text{ if } n+n' \text{ is even}\\
0 \text{ else}
\end{cases}\label{Mnl}
\end{equation}
with the coefficient\footnote{The change of sign of the $\alpha_{-}$ compared to Eq.~(A21) in \cite{devvisco}
is due to the convention $W_{E-}=-w_+$ and $W_{S-}=-w_-$ used in this manuscript.}
\begin{equation}
\alpha_{n\pm}^l=\frac{ 2}{2l+1} \frac{2\Omega_{S\pm}^l-(-1)^n\Omega_{E\pm}^l}{\Omega_{\Gamma}} \label{alphal}
\end{equation}
Here, $\Omega_\Gamma$ is the $l=0$ average of $W$ which enters in quasiparticle lifetime \eqqref{Gammaexplicite}:
\be
\Omega_\Gamma=\Omega_{E+}^0=\Omega_{S+}^0
\ee
%In the case of isotropic interactions ($W_{\upa\dwa}=\text{cte}$, $W_{\sigma\sigma}=0$), one has $\Omega_{S+}^l=\Omega_{\Gamma}$, $\Omega_{S-}^l=0$ and $\Omega_{E+}^l=-\Omega_{E-}^l=(-1)^l \Omega_{\Gamma}$.
Note that $\mathcal{M}^{0}_{\pm}\mathcal{F}^{\pm}_{0}=\mathcal{M}^0_{\pm}$ and $\mathcal{M}^{1}_+\mathcal{F}_1^{+} = \mathcal{M}^{1}_+ $ thanks to the conservation law of the collision integral.

\subsection{Numerical method}

We now present a numerical scheme to solve the projected transport equation \eqref{eq:transportprojBT} {based on a backward recurrence on $l$}. We
propagate backward in $l$ the linear relation between $\Vec{\nu}^l$ and $\Vec{\nu}^{l-1}$
\begin{equation}
    \Vec{\nu}^l = \mathcal{H}^l \Vec{\nu}^{l-1}
\end{equation}
%with $\mathcal{H}^l$ an infinite matrix. 
where we omit the $\pm$ index for convenience.  Numerically, we introduce a truncation parameter $n_{\rm max}$ and {represent the infinite matrix $\mathcal{H}^l$} by a complex $n_{\rm max} \times n_{\rm max}$ matrix. Substituting this relation into equation \eqref{eq:transportprojBT} for $l > 1$, we derive the backward recurrence relation on $\mathcal{H}^l$.
\begin{equation}
    \mathcal{H}^l = \frac{l}{2l-1}\left(\left(\frac{\omega}{\omega_0}+\frac{\ii}{\omega_0 \tau}\mathcal{M}^l\mathcal{F}_{l}\right)- \frac{l+1}{2l+3}\mathcal{F}_{l+1}\mathcal{H}^{l+1} \right)^{-1}\mathcal{F}_{l-1}
\end{equation}
{To initialize the recurrence,} we introduce a cutoff $l_{\rm max}$ and we assume $\mathcal{H}^{l_{\rm max}+1} = 0$.
At the end of the backward recurrence, we solve the remaining $2n_{\rm max}\times 2n_{\rm max}$ coupled system on $\Vec{\nu}^0$ and $\Vec{\nu}^1$.
For the density distribution $\nu^+$, this gives
\begin{equation}
\begin{cases}
\left( \frac{\omega}{\omega_0} + \frac{\ii}{\omega_0\tau}\mathcal{M}^0_+  \right)\Vec{\nu}^0_+ = \frac{1}{3}\mathcal{F}^+_1 \Vec{\nu}^1_+ \\
\left( \frac{\omega}{\omega_0} + \frac{\ii}{\omega_0\tau}\mathcal{M}^1_+  \right)\Vec{\nu}^1_+ - \mathcal{F}_0^+ \Vec{\nu}^0_+ -\frac{2}{5}\mathcal{F}_2^+ \mathcal{H}^2_+ \Vec{\nu}^1_+ = -\Vec{u}_0
\end{cases}
\end{equation}
Here, $\mathcal{H}^2_+$ is computed recursively, starting from $l = l_{\rm max}$. 
We choose the values of $l_{\rm max}$ and $n_{\rm max}$ based on a convergence analysis. Selecting cutoffs that are too low may lead to non-physical oscillations in the response functions.
In practice, $l_{\rm max}$ and $n_{\rm max}$ depend on the regime of $\omega_0 \tau$ under study.
In the collisionless regime, we can restrict ourselves to small values of $n_{\rm max}$.
This confirms the observation made in the previous section: the energy dependence of $\rho_{\pm}$ is contained in the collision term, which is subdominant.
Conversely, in the hydrodynamic regime, small values of $l_{\rm max}$ suffice, since the
non-conserved quantities with $l\geq 2$ decay as $(\omega_0\tau)^l$ \cite{disphydro}.

\subsection{Anisotropic driving potential for the polarisation}

The polarisation response to the isotropic drive introduced in \eqqref{Hext} vanishes as $\omega_0 \tau$ in the hydrodynamic regime. This is because such a drive couples to a dissipative component, the spin current $\nu^1_{0-}$. 
To make the diffusive mode of polarisation observable, one should rather couple the drive directly to the conserved quantity $n_{\uparrow}-n_{\downarrow}$, in the $l=0$ channel.
To do so, we assume that the driving potential can be varied independently with $\gq$ and $\grp$:
\begin{equation}\label{Hextpol}
    \hat{H}_{\rm ext} = \sum_{\grp \in \mathcal{D}}U_{-}(\grp,\gq) \left(\hat{\gamma}^{\dagger}_{\grp+\gq/2,\uparrow}\hat{\gamma}_{\grp-\gq/2,\uparrow} - \hat{\gamma}^{\dagger}_{\grp+\gq/2,\downarrow}\hat{\gamma}_{\grp-\gq/2,\downarrow}\right)
\end{equation}
The dependence of the driving potential on the modulus $p$ is irrelevant, so that we can write:
\begin{equation}
    U_{-}(\grp,\gq) = U_-(\gq) u(\theta)
\end{equation}
This change of $ \hat{H}_{\rm ext}$ modifies the source term in the polarisation transport equation:
\begin{equation}
    \left(\frac{\omega}{\omega_0}-\cos{\theta}\right)\nu_-(y,\theta) +\cos\theta\bbcro{u(\theta)- \frac{1}{2}\int \textrm{d}y'\frac{\textrm{d}\Omega'}{2\pi}F^-(\alpha)g(y')\nu_-(y',\theta')}  = - \textrm{i} I(y,\theta)
\end{equation}
To couple the drive directly to the  polarisation fluctuations, the product $u(\theta)\cos{\theta}$ should have a non-vanishing $l=0$ component, i.e. it should locally change
the spin imbalance. 
For simplicity, we omit here the components $l\geq 1$ whose contribution is negligible in the hydrodynamic limit, \textit{i.e.} we assume $u(\theta)\cos{\theta}=1$.
The set of equations to be solved at the end of the backward recurrence is then:
\begin{equation}
\begin{cases}
\left( \frac{\omega}{\omega_0} + \frac{\ii}{\omega_0\tau}\mathcal{M}^0_-  \right)\Vec{\nu}^0_- = \frac{1}{3}\mathcal{F}^-_1 \Vec{\nu} ^1_- -\Vec{u}_0 \\
\left( \frac{\omega}{\omega_0} + \frac{\ii}{\omega_0\tau}\mathcal{M}^1_- \mathcal{F}^-_1 \right)\Vec{\nu}^1_- - \mathcal{F}_0^- \Vec{\nu}^0_- -\frac{2}{5}\mathcal{F}_2^- \mathcal{H}^2_- \Vec{\nu}^1_- = 0
\end{cases}
\end{equation}

\subsection{Response functions in the collisionless-to-hydrodynamic crossover}

\begin{figure}[htb]
\centering

\includegraphics[width=0.6\textwidth]{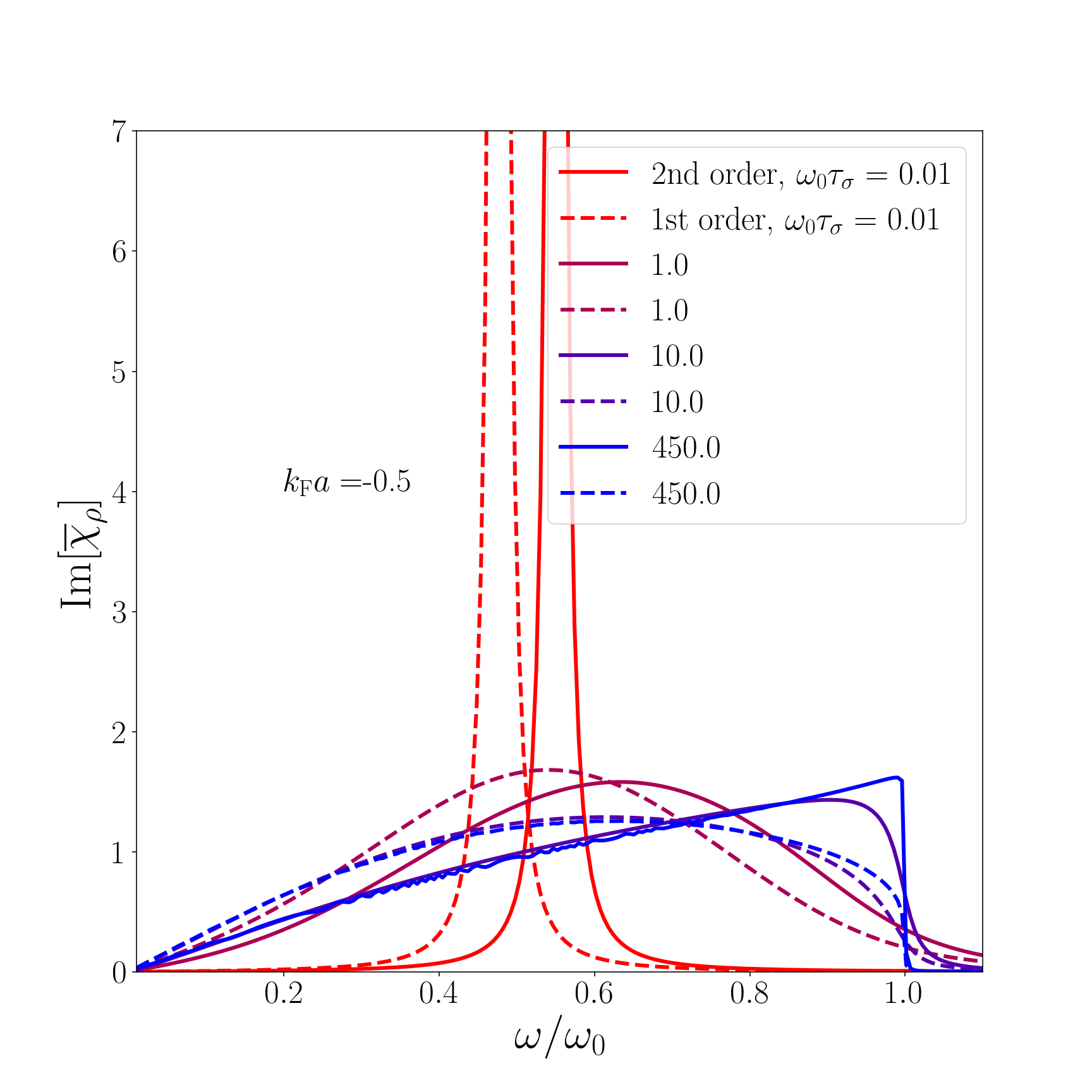}
\caption{The crossover between the hydrodynamic (red curves) and collisionless (blue curves) regimes in the reduced spectral density, $\textrm{Im}[\bar{\chi}_{\rho}] = \textrm{Im}[\nu^0_{0+}]$, at interaction strength $k_{\rm F}a = -0.5$. The collision parameter $\omega_0 \tau_{\sigma}$ is given by \eqref{tausigma}. Dashed lines indicate the solution with $f_{\sigma\sigma'}$ and $\mathcal{A}_{\sigma\sigma'}$ computed to leading order in $\kF a$, while solid lines include the second-order corrections. Here and in Figs.~\ref{fig:crossoverdensite0.5} and \ref{fig:crossoverpolarisation0.5}, the summation over $n$ is truncated from $n_{\rm max} = 50$ in the hydrodynamic regime to $n_{\rm max} = 5$ in the collisionless regime. Similarly, the truncation in $l$ is set to $l_{\rm max} = 5$ in the hydrodynamic regime and to $l_{\rm max} \approx \omega_0 \tau_{\sigma}$ outside of it.
\label{crossoverdensite-0.5}}
\end{figure}

\begin{figure}[htb]
\centering

\includegraphics[width=0.6\textwidth]{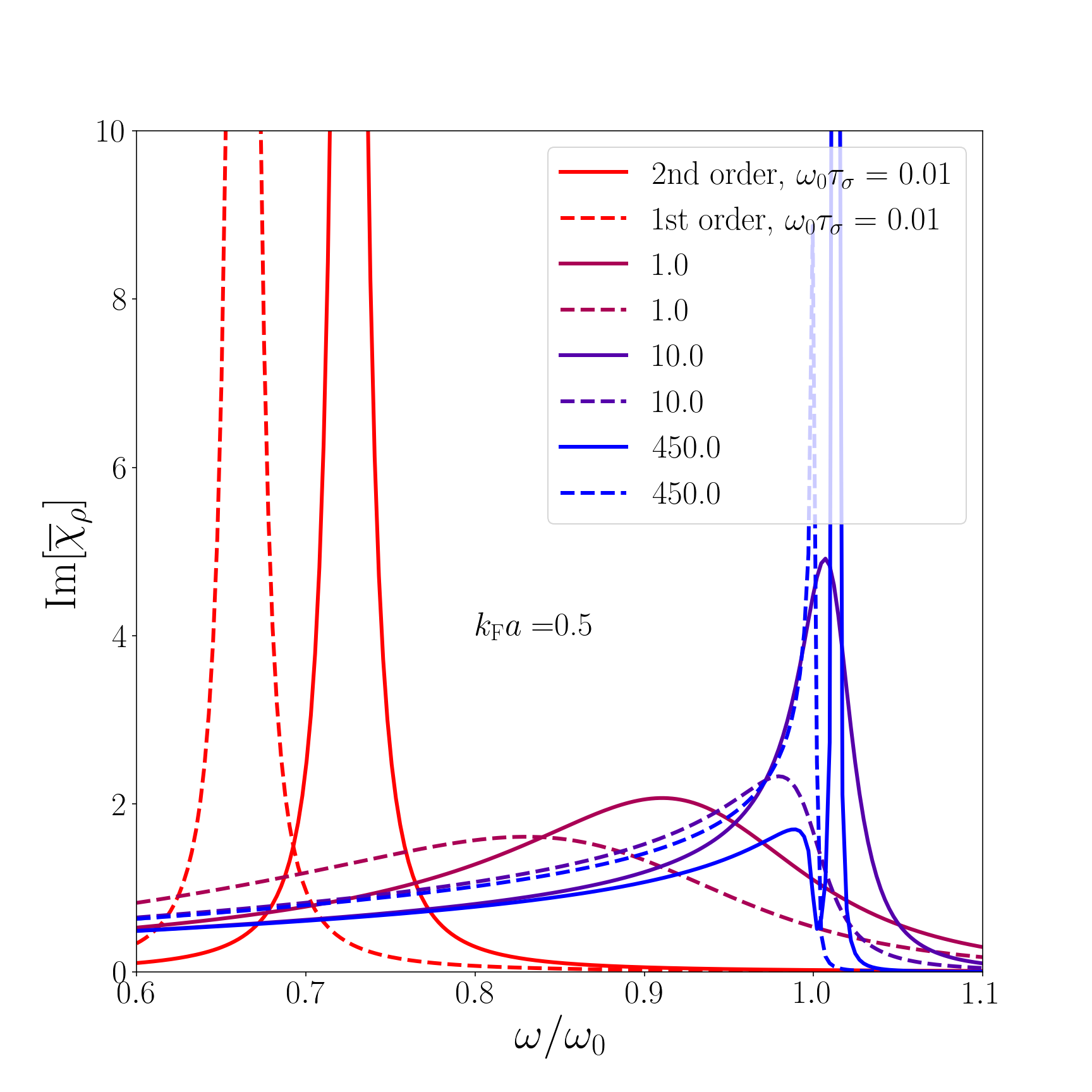}
\caption{The crossover between the hydrodynamic (red curves) and collisionless (blue curves) regimes in the reduced spectral density, $\textrm{Im}[\bar{\chi}_{\rho}] = \textrm{Im}[\nu^0_{0+}]$, at interaction strength $k_{\rm F}a = 0.5$. The collision parameter $\omega_0 \tau_{\sigma}$ is given by \eqref{tausigma}. Dashed lines indicate the first-order analytical solution, while solid lines correspond to the second-order correction.
\label{fig:crossoverdensite0.5}}
\end{figure}

\begin{figure}[htb]
\centering

\includegraphics[width=0.6\textwidth]{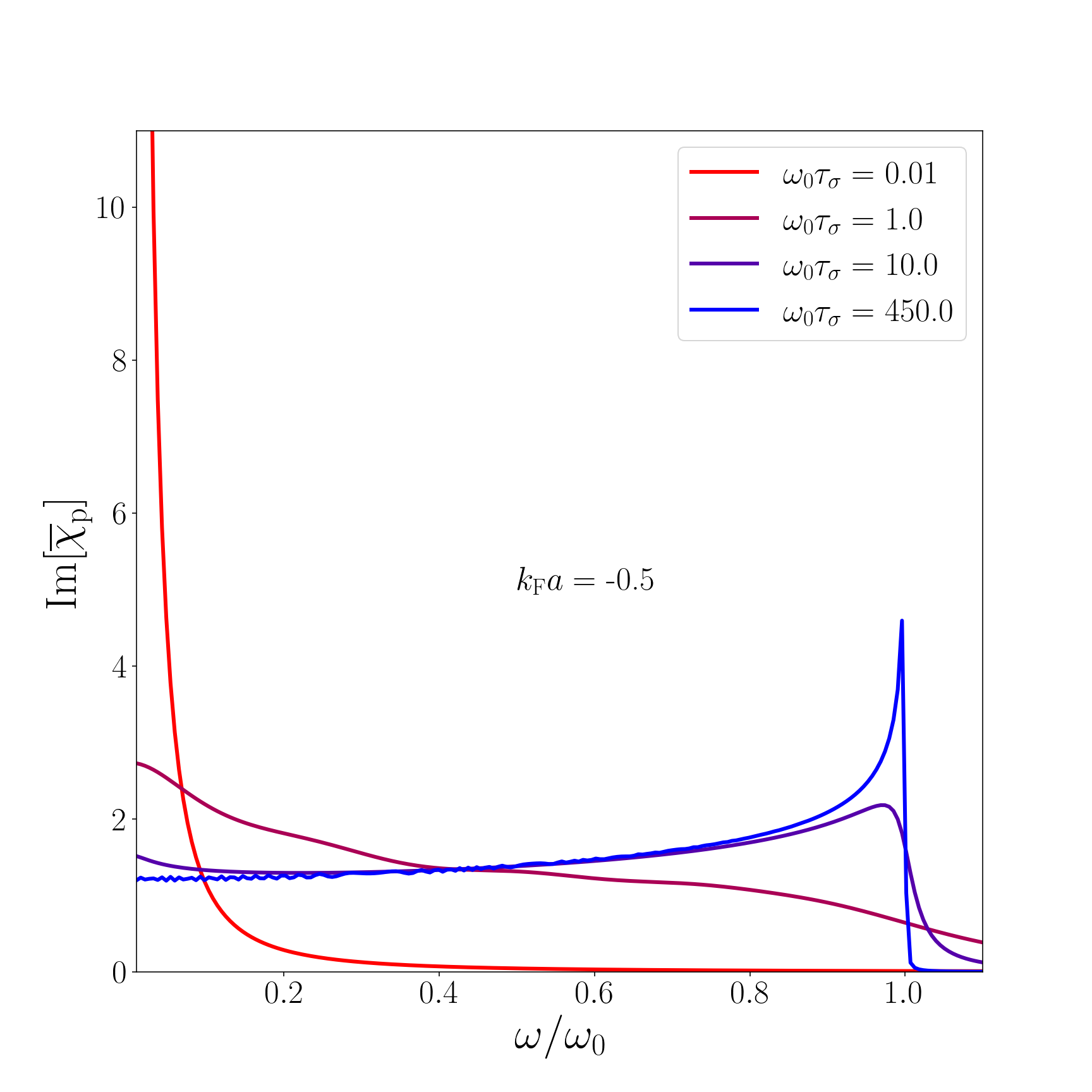}
\caption{The crossover between the hydrodynamic (red curves) and collisionless (blue curves) regimes in the reduced spectral density for the polarisation, $\textrm{Im}[\bar{\chi}_{\rm p}] = \textrm{Im}[\nu^0_{0-}]$, at interaction strength $k_{\rm F}a = -0.5$. The collision parameter $\omega_0 \tau_{\sigma}$ is given by \eqref{tausigma}.
\label{fig:crossoverpolarisation0.5}}
\end{figure}

We illustrate the collisionless-to-hydrodynamic crossover for the density (Figs.~\ref{crossoverdensite-0.5} and \ref{fig:crossoverdensite0.5} for the attractive and repulsive case respectively) and for the polarization response functions (Figs.~\ref{fig:crossoverpolarisation0.5}, for the attractive case).
As in \cite{devvisco}, we parametrize the crossover by $\omega_0\tau_{\sigma}$, where :
\begin{equation}\label{tausigma}
    \tau_{\sigma} = \frac{\pi}{2m\sigma T^2} \qquad \text{ with } \qquad \sigma = 4\pi a^2 
\end{equation}

We compare the first-order prediction (dashed curves) to the second-order prediction (solid curves) with the corrections to $f_{\sigma\sigma'}$ and $\mathcal{A}_{\sigma\sigma'}$ found in this work (Eqs.~(\ref{Aupdw}--\ref{fupup})).
In the density response function, we observe a shift of the first sound peak toward higher velocities when comparing the first-order and second-order calculations. Recall that the first sound velocity {(in units of $\vF$)} is given  by
\begin{equation}
c_1 = \sqrt{\frac{(1+F^+_0)(1+F^+_1/3)}{3}}.
\end{equation}
Since the {second-order} terms in $F_0^+$ and $F_1^+$ are positive {(irrespectively of the sign of $\kF a$)}, they increase the value of $c_1$ compared to the first-order result.

In the collisionless regime, a zero-sound mode is present in the repulsive case (Fig.~\ref{fig:crossoverdensite0.5}), possibly
with a secondary peak near the edge the continuum.
The resolution of those two peaks allows us to further divide the collisionless regime into two sub-regimes according to the value of $\omega_0 \tau_{\sigma}$.
For $1/(c_0-1) \gg \omega_0 \tau_{\sigma} \gg 1$ ($\omega_0 \tau_{\sigma} \approx 10$ in Fig.~\ref{fig:crossoverdensite0.5}) the zero-sound mode is not separated from the quasiparticle-hole continuum, which gives rise to a single peak
with an important left skewness.
A deeper collisionless regime, or true zero sound regime, is reached for $\omega_0 \tau_{\sigma} \gg 1/(c_0-1)$ ($\omega_0 \tau_{\sigma} \approx 450$ in Fig.~\ref{fig:crossoverdensite0.5}). In this regime, zero sound separates from the continuum, which however retains a significant spectral weight.
This deep regime is more easily reached when second order terms are included (compare  the blue solid and blue dashed curves in Fig.~\ref{fig:crossoverdensite0.5}); this is because $c_0 - 1$ is much larger in the second- than in the first-order approximation.
In between the hydrodynamic and collisionless regimes, the density response function retains a shallow maximum whose location smoothly evolves from $c_1$ to $c_0$. This peak is however too broad
to be identified as a collective mode: its width $\Delta c$ is comparable to 1 in units of $\vF$.

We now turn to the polarisation response (Fig~\ref{fig:crossoverpolarisation0.5}). In the collisionless regime (blue curves), we observe a skewed peaked at the continuum edge, but no zero sound resonance yet for $\omega_0 \tau_{\sigma}=450$ (blue curve).
Again, this can be understood by comparing $\omega_0 \tau_{\sigma}$ to $1/(c_0-1)$: the log-perturbative corrections from the second-order approximation reduce the deep collisionless regime  to $\omega_0 \tau_{\sigma}\gtrsim 10^5$ for $\kF a=-0.5$.
In the hydrodynamic regime (red curves), there appears a diffusive mode centered in $\omega=0$, as predicted by the Navier-Stokes equations of the Fermi liquid \cite{devvisco}.
Note that the large spectral weight of this peak is a consequence of our choice of an anisotropic drive \eqqref{Hextpol}. In between the collisionless and hydrodynamic limits, the polarisation response displays a very flat profile between
two local maxima in $c=0$ and $c\approx 1$.

\section*{Conclusion}
\label{sec:concl2}

Applying the effective picture to an atomic Fermi gas with contact interactions, we showed
how the use of Landau quasiparticles systematically improves the weak-coupling approximations.
In particular the celebrated Gor'kov-Melik Barkhudarov log-perturbative correction to $T_c$
and $\Delta$ emerges here as a direct manifestation of the quasiparticle dressing. 
Including second-order terms in the effective Hamiltonian, we also observe
large deviations  in the collisionless response functions compared to the RPA, 
including on the speed of zero sound $c_0$. The deviation $c_0-\vF$ is multiplied 
by $\text{exp}(6)$ for the density zero sound, and by $\text{exp}(-2)$ for the polarisation zero sound.

Extensions of this work could address the hydrodynamic 
regime where a normal quasiparticle fluid and a quasiparticle condensate coexist.
The Boltzmann and pairing equations derived here in the normal phase are a natural 
starting point for a microscopic derivation of the two-fluid 
hydrodynamics of Fermi systems \cite{Khalatnikov,Wolfle1990}.
More generally, the concept of Landau quasiparticles is not restricted
to unbalanced spin-$1/2$ Fermi systems, and applies more generally to quasiparticles whose
low-energy spectrum ressemble that of the free particle, as \textit{e.g.}
the Bose \cite{Bruun2018} and Fermi polarons \cite{Liu2018polaron}.
Our renormalization scheme could serve to derive an effective Hamiltonian
for such quasiparticles, including static interactions and collision amplitudes.
\section*{Acknowledgements}
Fruitful discussions with Nicolas Dupuis are gratefully acknowledged.
H.K. acknowledges support from the French Agence Nationale de la Recherche (ANR), under grant ANR-23-ERCS-0005 (project DYFERCO).

\begin{appendix}
\renewcommand{\theequation}{II.\Alph{section}\arabic{equation}}
\section{$\Lambda$ dependence of the collision amplitudes}
\label{app:ijlambda}
In this appendix, we detail the calculation of the functions
$I_\Lambda$ and $J_\Lambda$ introduced in Sec.~\ref{sec:lambdadep} (see also Figs.~\ref{ilambda} and \ref{jlambda})
to characterize the angular dependence of $\mathcal{B}_{\sigma\sigma'}$.
Comparing \eqqrefs{Bpss}{Bpud} and \eqqrefs{Bupdw}{Bupup}, we identify the dimensionless
coefficients of the $O(\kF a)^2$ terms in $\mathcal{B}_{\sigma\sigma'}$:
\begin{align}
I_\Lambda(\pp,\pp')&=\frac{(2\pi)^2\EF}{(\pF L)^3}
\sum_{\pp_1\pp_2\in\mathcal{D}}\bbcro{n_{\pp_1}^0+ n_{\pp_2}^0}\delta_{\pp+\pp'}^{\pp_1+\pp_2}{\mathcal{P}_{\Lambda}\bb{\frac{1}{\omega_{\pp_1}+\omega_{\pp_2}-2\EF}}} \\
J_\Lambda(\pp,\pp')&=-\frac{(2\pi)^2\EF}{(\pF L)^3}\sum_{\pp_1\pp_2\in\mathcal{D}}\bbcro{n_{\pp_1}^0 - n_{\pp_2}^0}\delta_{\pp_1+\pp'}^{\pp_2+\pp}{\mathcal{P}_{\Lambda}\bb{\frac{1}{\omega_{\pp_1}-\omega_{\pp_2}}}}
\end{align}
With $p=p'=\pF$, the functions depend on $\pp$ and $\pp'$ only through the angle $\alpha=(\widehat{\pp,\pp'})$. We eliminate $\pp_2$ through momentum conservation and locate $\pp_1$ in a spherical frame with $\pp+\pp'$ or $\pp-\pp'$ as the $z$-axis respectively in $I_\Lambda$ and $J_\Lambda$. 
Exploiting the invariance on the azimuthal angle, parameterizing the polar angle by $u=\cos\theta_1$, and introducing the dimensionless momentum $x=p_1/\pF$, we write
\begin{align}
I_\Lambda(\theta)&=-2\int_0^1 x^2 \dd x \int_{-1}^1 \dd u \mathcal{P}_{\epsilon}\bb{\frac{1}{2(2cx u-x^2-2c^2+1)}}\\
J_\Lambda(\theta)&=-2\int_0^1 x^2 \dd x \int_{-1}^1 \dd u  \mathcal{P}_{\epsilon}\bb{\frac{1}{4(x s u-s^2)}}
\end{align}
where $\mathcal{P}_{\epsilon}(1/f)=\Theta(|f|-\epsilon)/f$ is the $\epsilon$–regularized principal part. We parametrize the
small parameter associated to $\Lambda$ using
\be
\epsilon=\frac{\Lambda}{4\EF}, \qquad \epsilon'=2\epsilon
\ee
where $\epsilon$ coincides with the $\epsilon_\Lambda$ used in the main text.
We have also parametrized the $\alpha$-dependence through
\begin{align}
c&=\cos\frac{\alpha}{2}=\frac{||\pp+\pp'||}{2\pF} \\
s&=\sin\frac{\alpha}{2}=\frac{||\pp-\pp'||}{2\pF}
\end{align}
The $\epsilon$-principal part excludes a region of the integration domain $[0,1]\times[-1,1]$, and this forbidden band varies with $\alpha$ and $\epsilon$.
\begin{figure}
\begin{center}
\includegraphics[width=0.9\textwidth]{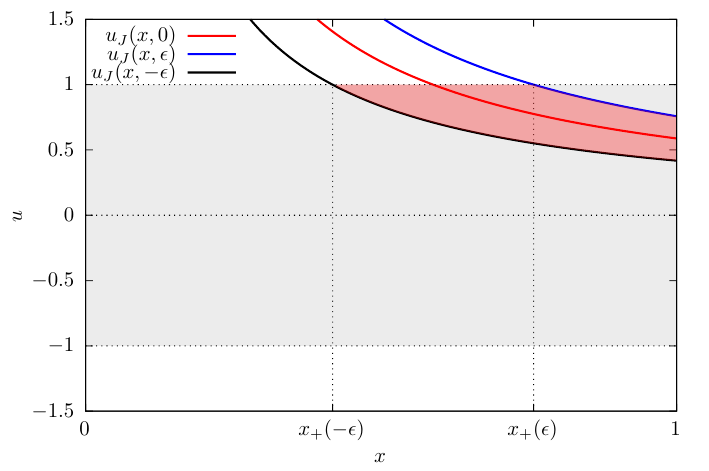}
\end{center}
\caption{\label{fig:res} The resonance angle $u_J(x,\epsilon=0)$ (red curve) and the forbidden band $x\mapsto[u_J(x,-\epsilon),u_J(x,\epsilon)]$ (red area) inside the integration domain $[0,1]\times[-1,1]$ (grey area) for $J_\Lambda$ at $\alpha=0.4\pi$ and $\epsilon=0.1$.}
\end{figure}
To identify the excluded region in the integration interval $[-1,1]$ over $u$, we introduce the resonance angles
\begin{align}
u_I(x,\epsilon')&=\frac{x^2+2c^2-1+\epsilon'}{2cx}\\
u_J(x,\epsilon)&=\frac{s^2+\epsilon}{sx}
\end{align}
The $\epsilon$-resonance conditions then read $u_{I}(x,-\epsilon')\leq u\leq u_{I}(x,\epsilon')$ and $u_{J}(x,-\epsilon)\leq u\leq u_{J}(x,\epsilon)$,
which allows to rewrite $I$ and $J$ as:
\begin{align}
I_\Lambda(\theta)&=-\int_0^1 \frac{x \dd x}{2c} \int_{-1}^1 \frac{\dd u}{u-u_I(x,0)}\bb{1-\Theta\bbcro{u_I(x,\epsilon')-u}\Theta\bbcro{u-u_I(x,-\epsilon')}} \\
J_\Lambda(\theta)&=-\int_0^1 \frac{x\dd x}{2s}  \int_{-1}^1 \frac{\dd u}{u-u_J(x,0)} \bb{1-\Theta\bbcro{u_J(x,\epsilon)-u}\Theta\bbcro{u-u_J(x,-\epsilon)}}
\end{align}
Fig.~\ref{fig:res} shows an example of the forbidden band in the calculation of $J_\Lambda$ at $\alpha=0.4\pi$ and $\epsilon=0.1$.

\paragraph{Expression of $I_\Lambda$} Depending on the comparison of $u_{I}(x,\pm\epsilon')$ with $\pm1$, the excluded band may be
$\emptyset$, the interval $[u_{I}(x,-\epsilon'),u_{I}(x,\epsilon')]$, $[u_{I}(x,-\epsilon'),1]$, $[-1,u_{I}(x,\epsilon'),1]$ or $[-1,1]$.
Upon integration over $u$, this generates $3$ different integrands of $x$:
\begin{alignat}{3}
&f(x)&=&-\frac{x}{2c}\int_{-1}^{1}\frac{\dd u}{u-u_I(x,0)}&=\frac{x}{2c}\text{ln}\left\vert\frac{(c+s+x)(c-s+x)}{(c+s-x)(c-s-x)}\right\vert\\
&f_{+\epsilon'}(x)&=&-\frac{x}{2c}\int_{u_I(x,\epsilon')}^{1}\frac{\dd u}{u-u_I(x,0)}&=\frac{x}{2c}\text{ln}\left\vert\frac{\epsilon'}{(c+s-x)(c-s-x)}\right\vert\\
&f_{-\epsilon'}(x)&= &-\frac{x}{2c}\int_{-1}^{u_I(x,-\epsilon')}\frac{\dd u}{u-u_I(x,0)}&=\frac{x}{2c}\text{ln}\left\vert\frac{(c+s+x)(c-s+x)}{\epsilon'}\right\vert
\end{alignat}
Note that $f$ also describes the integral for $u\in[-1,u_I(x,-\epsilon')]\cup [u_I(x,\epsilon'),1]$.
The remaining integral over $x$ is divided in up to 4 intervals, where either one
of the functions $f$, $f_{+\epsilon'}$ or $f_{-\epsilon'}$ is used. The bounds delimiting these intervals (see Fig.~\ref{fig:res}) are
\be
x_\pm(\epsilon')=c\pm\sqrt{s^2-\epsilon'}
\ee
Again, depending on $\alpha$, the boundaries $x_\pm(\pm\epsilon')$ maybe inside or outside
the integration interval $[0,1]$ over $x$. This generates 8 different slicing configurations
of $[0,1]$, listed in function of $\alpha$ on Fig.~\ref{fig:configI}. 
The corresponding expression of $I_\Lambda$ is given in \eqqrefs{iIII}{iIIIp}.
We stitch together these expressions over the domain of variation $[0,\pi]$ of $\alpha$
to produce the red curve in Fig.~\ref{ilambda} of the main text.
\begin{figure}
\begin{tikzpicture}[xscale=0.85]
\draw[->] (0,0) -- (16,0);
\draw[->] (0,-2) -- (16,-2);
\draw (15.2,-2) node[below right=0.0cm] {$c=\cos\frac{\alpha}{2}$};
\draw (16,0) node[above=0.1cm] {$\alpha$};
\draw (1,0) node{$\vert$} node[above=0.2cm] {$0$} -- (1,-2);
\draw (1,-2) node{$\vert$} node[below=0.2cm] {$1$};
\draw (1.5,-1) node {III};
\draw (2,0) node{$\vert$} node[above=0.2cm] {$\alpha_1$}  -- (2,-2);
\draw (2,-2) node{$\vert$} node[below=0.2cm] {$\sqrt{1-\epsilon'}$};
\draw (3,-1) node {II};
\draw (4,0) node{$\vert$} node[above=0.2cm] {$\alpha_2$} -- (4,-2);
\draw (4,-2) node{$\vert$} node[below=0.2cm] {$\frac{1+\sqrt{1-2\epsilon'}}{2}$};
\draw (5,-1) node {I};
\draw (6,0) node{$\vert$} node[above=0.2cm] {$\alpha_{t_1}$}  -- (6,-2);
\draw (6,-2) node{$\vert$} node[below=0.2cm] {$\sqrt{\frac{1+\epsilon'}{2}}$};
\draw (7,-1) node {A};
\draw (8,0) node{$\vert$} node[above=0.2cm] {$\alpha_{t_2}$}  -- (8,-2);
\draw (8,-2) node{$\vert$} node[below=0.2cm] {\scriptsize $\frac{1}{2}\sqrt{1+\sqrt{1-2\epsilon'^2}}$};
\draw (9,-1) node {B};
\draw (10,0) node{$\vert$} node[above=0.2cm] {$\alpha_{t_3}$}  -- (10,-2);
\draw (10,-2) node{$\vert$} node[below=0.2cm] {$\sqrt{\frac{1-\epsilon'}{2}}$};
\draw (11,-1) node {I'};
\draw (12,0) node{$\vert$} node[above=0.2cm] {$\alpha_{2}'$}  -- (12,-2);
\draw (12,-2) node{$\vert$} node[below=0.2cm] {$\scriptsize\frac{1-\sqrt{1-2\epsilon'}}{2}$};
\draw (13,-1) node {II'};
\draw (14,0) node{$\vert$} node[above=0.2cm] {$\alpha_{1}'$}  -- (14,-2);
\draw (14,-2) node{$\vert$} node[below=0.2cm] {$\scriptsize\frac{\sqrt{1-2\epsilon'}-1}{2}$};
\draw (14.5,-1) node {III'};
\draw (15,0) node{$\vert$} node[above=0.2cm] {$\pi$}  -- (15,-2);
\draw (15,-2) node{$\vert$} node[below=0.2cm] {$0$};
\end{tikzpicture}
\caption{\label{fig:configI} As $\alpha$ varies from $0$ to $\pi$, $I_\Lambda$ assumes 8 different expressions
(\eqqrefs{iIII}{iIIIp}). The corner points $\alpha_n$ that separate these expressions are given by $\cos\frac{\alpha_n}{2}=i_n(\epsilon')$, where $i_n(\epsilon')$ is given on the lower axis.}
\end{figure}
\begin{align}
I_{\Lambda}^{\rm III}&=\int_{0}^{x_-(-\epsilon')} {\dd x} f(x) +\int_{x_-(-\epsilon')}^{1}\dd xf_{-\epsilon'}(x) \label{iIII} \\
I_{\Lambda}^{\rm II}&=\int_{0}^{x_-(-\epsilon')} {\dd x} f(x) +\int_{x_-(-\epsilon')}^{x_-(\epsilon')}\dd xf_{-\epsilon'}(x) +\int_{x_-(\epsilon')}^{x_+(\epsilon')} \dd x f(x) +\int_{x_+(\epsilon')}^{1} \dd x f_{-\epsilon'}(x) \\
I_{\Lambda}^{\rm I}&=\int_{0}^{x_-(-\epsilon')} {\dd x} f(x) +\int_{x_-(-\epsilon')}^{x_-(\epsilon')}\dd xf_{-\epsilon'}(x) +\int_{x_-(\epsilon')}^1 \dd x f(x) \\
I_{\Lambda}^{\rm A}&=\int_{|x_-(-\epsilon')|}^{x_-(\epsilon')} {\dd x} f_{-\epsilon'}(x) +\int_{x_-(\epsilon')}^{1}\dd xf(x)\\
I_{\Lambda}^{\rm B}&=\int_{x_-(\epsilon')}^{|x_-(-\epsilon')|} {\dd x} f_{+\epsilon'}(x) +\int_{|x_-(-\epsilon')|}^{1}\dd xf(x)\\
I_{\Lambda}^{\rm I'}&=\int_{0}^{|x_-(\epsilon')|} {\dd x} f(x) +\int_{|x_-(\epsilon')|}^{|x_-(-\epsilon')|}\dd xf_{+\epsilon'}(x) +\int_{|x_-(-\epsilon')|}^1 \dd x f(x) \\
I_{\Lambda}^{\rm II'}&=\int_{0}^{|x_-(\epsilon')|} {\dd x} f(x) +\int_{|x_-(\epsilon')|}^{|x_-(-\epsilon')|}\dd xf_{+\epsilon'}(x) +\int_{|x_-(-\epsilon')|}^{x_+(\epsilon')} \dd x f(x)+\int_{x_+(\epsilon')}^1 \dd x f_{-\epsilon'}(x)\\
I_{\Lambda}^{\rm III'}&=\int_{0}^{|x_-(\epsilon')|} {\dd x} f(x) +\int_{|x_-(\epsilon')|}^{x_+(-\epsilon')}\dd xf_{+\epsilon'}(x) \label{iIIIp}
\end{align}

\paragraph{Expression of $J_\Lambda$} Similarly, for $J_\Lambda$, 
the excluded band in $u$ is either $\emptyset$, $[u_{J}(x,-\epsilon),u_{J}(x,\epsilon)]$, $[u_{J}(x,-\epsilon),1]$ or $[-1,1]$, and the corresponding integrands are
\begin{alignat}{3}
&g(x)&=&-\frac{x}{2s}\int_{-1}^{1}\frac{\dd u}{u-u_J(x,0)}&=\frac{x}{2s}\text{ln}\left\vert\frac{x+s}{x-s}\right\vert\\
&g_{-\epsilon}(x)&= &-\frac{x}{2s}\int_{-1}^{u_J(x,-\epsilon)}\frac{\dd u}{u-u_J(x,0)}&=\frac{x}{2s}\text{ln}\left\vert\frac{x+s}{2\epsilon}\right\vert
\end{alignat}
The interval [0,1] of integration over $x$ is divided by the boundaries
\be
x_\pm(\epsilon)=\pm\bb{s+\frac{\epsilon}{s}}
\ee
into 5 possible configurations listed in Fig.~\ref{fig:configJ}.
\begin{figure}
\begin{tikzpicture}[xscale=0.85]
\draw[->] (0,0) -- (16,0);
\draw[->] (0,-2) -- (16,-2);
\draw (15.2,-2) node[below right=0.0cm] {$s=\sin\frac{\alpha}{2}$};
\draw (16,0) node[above=0.1cm] {$\alpha$};
\draw (1,0) node{$\vert$} node[above=0.2cm] {$0$} -- (1,-2);
\draw (1,-2) node{$\vert$} node[below=0.2cm] {$0$};
\draw (2.5,-1) node {I};
\draw (4,0) node{$\vert$} node[above=0.2cm] {$\alpha_1$}  -- (4,-2);
\draw (4,-2) node{$\vert$} node[below=0.2cm] {$\frac{\sqrt{1+4\epsilon}-1}{2}$};
\draw (5.5,-1) node {II};
\draw (7,0) node{$\vert$} node[above=0.2cm] {$\alpha_2$} -- (7,-2);
\draw (7,-2) node{$\vert$} node[below=0.2cm] {$\frac{1-\sqrt{1-4\epsilon}}{2}$};
\draw (8.5,-1) node {III};
\draw (10,0) node{$\vert$} node[above=0.2cm] {$\alpha_{3}$}  -- (10,-2);
\draw (10,-2) node{$\vert$} node[below=0.2cm] {${{\sqrt{\epsilon}}}$};
\draw (11.5,-1) node {IV};
\draw (13,0) node{$\vert$} node[above=0.2cm] {$\alpha_{4}$}  -- (13,-2);
\draw (13,-2) node{$\vert$} node[below=0.2cm] {$\frac{1+\sqrt{1-4\epsilon}}{2}$};
\draw (14,-1) node {V};
\draw (15,0) node{$\vert$} node[above=0.2cm] {$\pi$}  -- (15,-2);
\draw (15,-2) node{$\vert$} node[below=0.2cm] {$1$};
\end{tikzpicture}
\caption{\label{fig:configJ} As $\alpha$ varies from $0$ to $\pi$, $J_\Lambda$ assumes 5 different expressions
(\eqqrefs{jI}{jV}). The corner points $\alpha_n$ that separate these expressions are given by $\sin\frac{\alpha_n}{2}=j_n(\epsilon)$, where $j_n(\epsilon)$ is given on the lower axis.}
\end{figure}
The corresponding expressions of $J_\Lambda$ are
\begin{align}
J_{\Lambda}^{\rm I}&=0 \label{jI}\\
J_{\Lambda}^{\rm II}&=\int_{x_-(-\epsilon)}^1 {\dd x} g_{-\epsilon}(x)  \\
J_{\Lambda}^{\rm III}&=\int_{x_-(-\epsilon)}^{x_+(\epsilon)} {\dd x} g_{-\epsilon}(x) +\int_{x_+(\epsilon)}^1 {\dd x} g(x)  \\
J_{\Lambda}^{\rm IV}&=\int_{0}^{x_+(-\epsilon)} {\dd x} g(x)+\int_{x_+(-\epsilon)}^{x_+(\epsilon)} {\dd x} g_{-\epsilon}(x) +\int_{x_+(\epsilon)}^1 {\dd x} g(x)  \\
J_{\Lambda}^{\rm V}&=\int_{0}^{x_+(-\epsilon)} {\dd x} g(x)+\int_{x_+(-\epsilon)}^{1} {\dd x} g_{-\epsilon}(x)  \label{jV} 
\end{align}
Combined as prescribed by Fig.~\ref{fig:configI}, these expressions produce the red curve in Fig.~\ref{jlambda} of the main text.
Note that the $\epsilon=0$ expressions of $I$ and $J$ (\eqqrefs{itheta}{jtheta}) are given by
\begin{align}
I(\alpha)&=\int_0^1\dd x f(x)\\
J(\alpha)&=\int_0^1\dd x g(x)
\end{align}

\section{Numerical evaluation of the zero sound velocity}\label{app:sonzero}

We present in this appendix the numerical method used to solve \eqref{transportCollproj}, and benchmark the analytic solution \eqref{gammadispersioncoll} of the prefactor $\exp(\gamma_1^\pm)$ in $c_0^{\pm}$

Recall that the transport equation for $\nu^{\pm}$ in the collisionless limit projects as:
\begin{equation}
    \nu^{\pm}_l(c) - \sum_{l'} A_{ll'}^{\pm}(c)\nu^{\pm}_{l'}(c) + B_{l0}(c) = 0 
\end{equation}
As mentioned in the main text, to compute $c_0$, we look for the zeros of the following determinant:
\begin{equation}\label{eq:solzero}
    \mathrm{Det}\left(1-\mathcal{A}_{\pm}(c_0^{\pm})\right) = 0 
\end{equation}
To do so, we truncate the matrix $\mathcal{A}_{\pm}$ at some $l_{\rm max}$, and we check the convergence with respect to this parameter. Typically, $l_{\rm max}\approx50$ is sufficient. 
To overcome the numerical limitation to $|k_{\rm F}a|>~0.1$, we perform a second-order polynomial extrapolation $\gamma^\pm+2-\pi/\ab = A+B\ab+C\ab^2$.

We find that $\gamma_{1}^{\pm} = \pm 4$ within the numerical accuracy of our extrapolation. The coefficient $B$ obtained from the extrapolation of $\gamma^{\pm}_1$ is larger than one, which restrict the observability of $\gamma_1^{\pm}$ to $|\ab|<0.1$.% that is $c_0-1 \approx 10^{-?}$.

We present in Figs.~\ref{prefacteurmoins} and \ref{prefacteurplus} these numerical interpolations.

\begin{figure}
    \centering
    \includegraphics[width=0.6\textwidth]{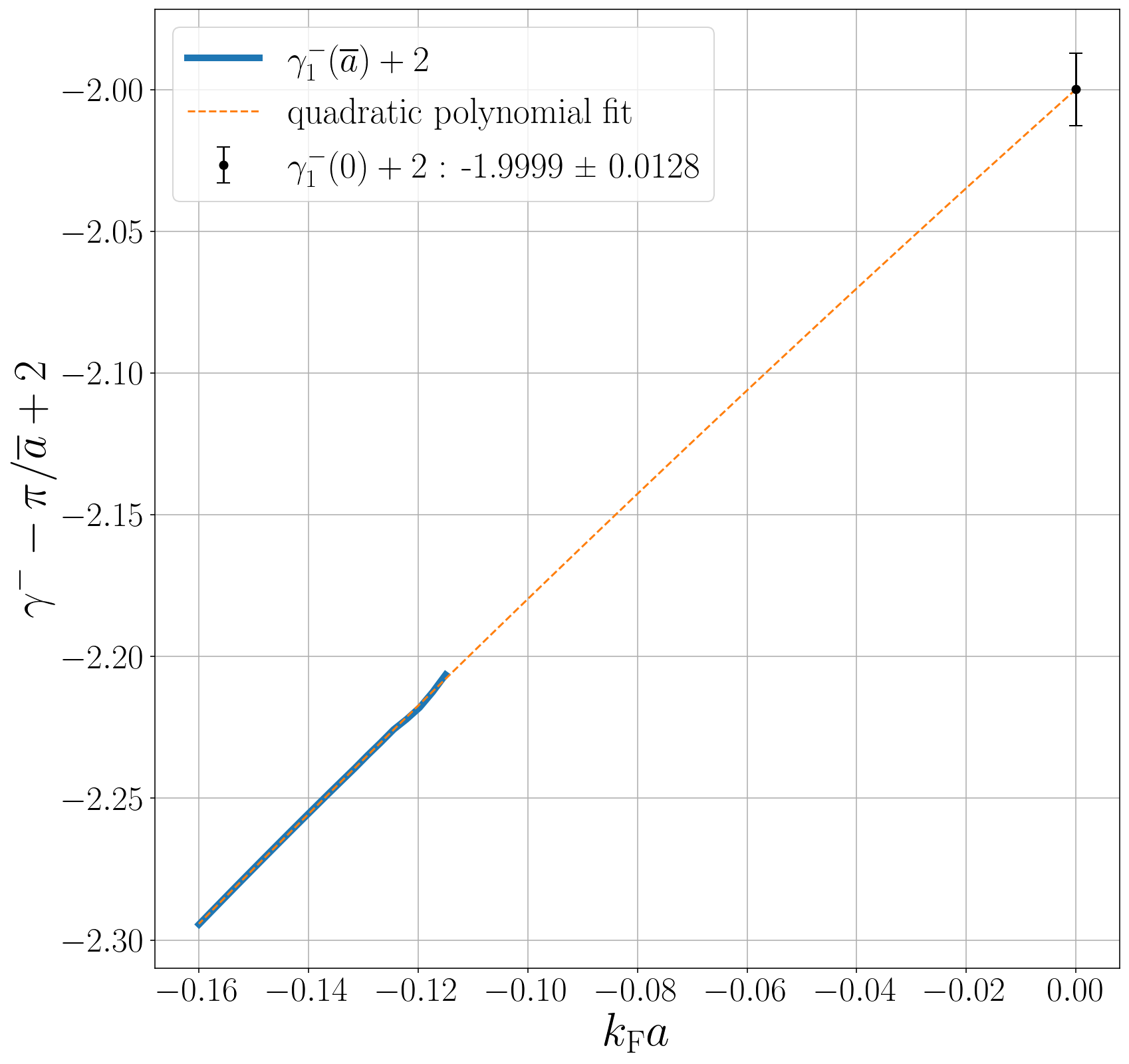}
    \caption{The reduced speed of the collisionless polarisation sound $\gamma^- +2 -\pi/ \ab $ with $\gamma^-=\log{((c_0^--1)/2)}$. The blue curve is obtained by numerically solving \eqref{eq:solzero} in the range $k_{\rm F}a=-0.16, -0.125$. A quadratic polynomial fit (orange curve) provides the value extrapolated to $k_{\rm F}a=0$: $\gamma_1^- +2= -1.9999 \pm 0.0128$.\label{prefacteurmoins}}
\end{figure}

\begin{figure}
    \centering
    \includegraphics[width=0.6\textwidth]{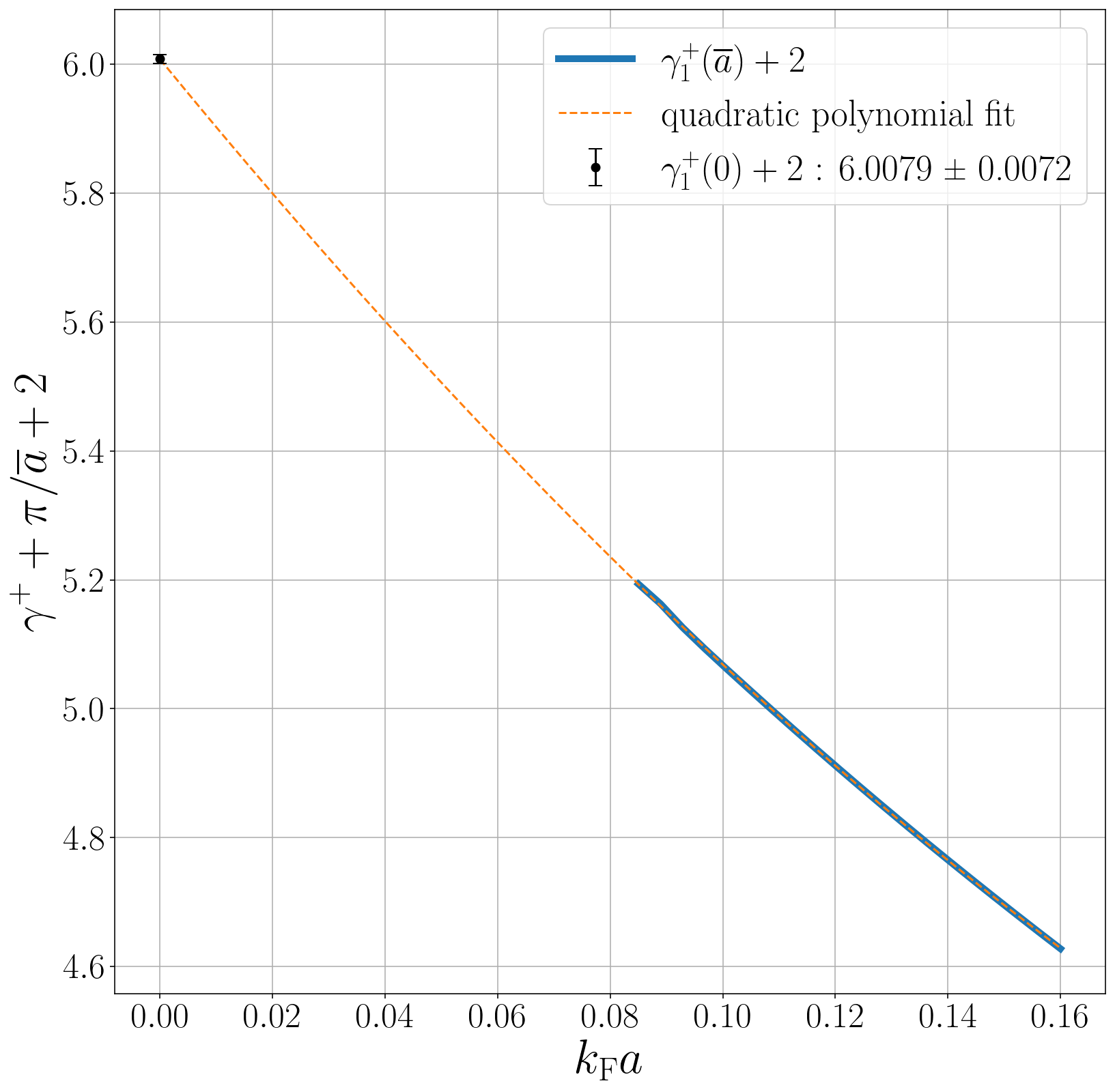}
    \caption{The reduced speed of the collisionless density sound $\gamma^+ +2 +\pi/ \ab $ with $\gamma^+=\log{((c_0^+-1)/2)}$. The blue curve is obtained by numerically solving \eqref{eq:solzero} in the range $k_{\rm F}a=0.16, 0.085$. A quadratic polynomial fit (orange curve) provides the value extrapolated to $k_{\rm F}a=0$: $\gamma_1^+ +2= 6.0079 \pm 0.0072$.\label{prefacteurplus}}
\end{figure}

\section{Collision effects in the collisionless regime}\label{app:colldamp}
In this appendix we present the calculation of the function $C_{\pm}$ introduced in \eqref{CollisionCollessKernel}:
\begin{equation}
    C_{\pm}(c) = \int^{\pi}_0 \mathrm{d}\theta \ \frac{\sin{\theta}\cos{\theta}}{(c-\cos{\theta})^2} + \int^{2\pi}_0 \frac{\mathrm{d}\phi'}{2\pi}\int^{\pi}_0 \mathrm{d}\theta\mathrm{d}\theta' \ \frac{\sin{\theta}\sin{\theta'}\cos{\theta'}}{(c-\cos{\theta})(c-\cos{\theta'})}\mathcal{N}_{\pm}(\alpha)
\end{equation}
with the angular collision kernel and its projection on Legendre polynomials given by:
\begin{equation}
    \mathcal{N}_{\pm}(\alpha)  = \frac{\Omega_{E\pm}(\alpha)-2\Omega_{S\pm}(\alpha)}{\Omega_{\Gamma}} = \sum_{l}\mathcal{N}_{\pm}^l(\alpha) P_l(\cos{\alpha})
\end{equation}
We use the addition theorem:
\begin{equation}
    \int^{2\pi}_0 \frac{\mathrm{d}\phi'}{2\pi} P_l(\cos\alpha) = P_l(\cos{\theta})P_l(\cos{\theta'})
\end{equation}
This allows us to factorize the integrals over $u=\cos\theta$ and $u'=\cos\theta'$ in $C_{\pm}$:
\begin{equation}
    C_{\pm}(c) = \int^{1}_{-1} \mathrm{d}u \ \frac{u}{(c-u)^2} + \sum_l \mathcal{N}_{\pm}^l \int^{1}_{-1} \mathrm{d}u \ \frac{P_{l}(u)}{c-u} \ \int^{1}_{-1} \mathrm{d}u' \ \frac{u' P_{l}(u')}{c-u'} 
\end{equation}
The different integrals are given by:
\begin{equation}
    \int^{1}_{-1} \mathrm{d}u \ \frac{u}{(c-u)^2} = - B_{00}'(c)
\end{equation}

\begin{equation}
    \int^{1}_{-1} \mathrm{d}u \ \frac{P_{l}(u)}{c-u} = 2 R_l(c)
\end{equation}

\begin{equation}
    \int^{1}_{-1} \mathrm{d}u' \ \frac{u' P_{l}(u')}{c-u'} = 2 c R_l(c) - 2\delta_{l,0}
\end{equation}
where we have introduced the Legendre functions of the second kind $R_l$ \cite{Gradshteyn}. The contribution of collisions is therefore finally contained in the following formula:
\begin{equation}
    C_{\pm}(c) = -B_{00}'(c) + 4 \sum_l \mathcal{N}^l_{\pm} R_l(c) (cR_l(c)-\delta_{l,0})
\end{equation}
In fact, this last formula is quite general, as the characteristics of the interactions are contained in the collision parameter $\mathcal{N}_{\pm}^l$.
We can now focus on the asymptotic behavior when $c$ tends exponentially to $1$.
Note that:
\begin{equation}
    \frac{R_0(c)(c R_0(c)-1)}{B_{00}'(c)} \underset{\gamma\to-\infty}{\sim} -\gamma^2 \textrm{e}^{\gamma}   
\end{equation}
Similarly, for $l>0$:
\begin{equation}
    \frac{R_l^2(c)}{B_{00}'(c)} \underset{\gamma\to-\infty}{\sim} -\gamma^2 \textrm{e}^{\gamma}  
\end{equation}
In all cases, we can rewrite the function $C_{\pm}$ in the limit where $c$ tends exponentially to $1$ as:
\begin{equation}
    C_{\pm}(c) \simeq -B_{00}'(c) \left(1+4{\gamma^2\text{e}^{\gamma}} \ \mathcal{N}_{\pm}(0)\right)
\end{equation}
where we have recognized that $\sum_l \mathcal{N}_{\pm}^l = \mathcal{N}_{\pm}(0)$.

\end{appendix}

\bibliography{biblio.bib}

\end{document}